\documentclass[review,10pt]{elsarticle}
\usepackage[a4paper,
    top=4.3cm,
    bottom=4.3cm,
    left=4.8cm,
    right=4.8cm]{geometry}

\usepackage{amssymb}
\usepackage{amsmath}

\usepackage{stix} % added by KWO
\usepackage{multirow} % added by KWO
\usepackage[automake]{glossaries} % added by KWO
\usepackage{xcolor} % added by KWO
\usepackage{hyperref} % added by KWO
\usepackage[utf8]{inputenc} % added by KWO
\usepackage[T1]{fontenc} % added by KWO
\usepackage{geometry} % added by KWO
\usepackage{graphicx} % added by KWO
\usepackage{subcaption} % added by KWO
\usepackage{float} % added by KWO
\usepackage{adjustbox} % added by KWO
\usepackage[dvipsnames]{xcolor} % added by KWO
\usepackage{tikz} % added by KWO
\usetikzlibrary{shapes.geometric, arrows} % added by KWO
\usepackage{utfsym} % added by KWO
\usetikzlibrary{mindmap,trees} % added by KWO
\usetikzlibrary{backgrounds,positioning} % added by KWO
\usetikzlibrary{mindmap,backgrounds} % added by KWO
\usetikzlibrary{decorations.markings} % added by KWO
\usetikzlibrary{positioning} % added by KWO
\usetikzlibrary{calc} % added by KWO

\newcommand\independent{\protect\mathpalette{\protect\independenT}{\perp}}
\def\independenT#1#2{\mathrel{\rlap{$#1#2$}\mkern2mu{#1#2}}}

\journal{Pattern Recognition}

\makeglossaries
\newacronym{aic}{AIC}{Akaike Information Criterion}
\newacronym{ar1}{AR(1)}{Autoregressive models of order one}
\newacronym{bf}{BF}{Bayes Factor}
\newacronym{bic}{BIC}{Bayesian Information Criterion}
\newacronym{bydv}{BYDV}{Barley Yellow Dwarf Virus}
\newacronym{glm}{GLM}{Generalised Linear Models}
\newacronym{gp}{GP}{Gaussian Processes}
\newacronym{grf}{GRF}{Gaussian Random Field}
\newacronym{gwr}{GWR}{Geographically Weighted Regression}
\newacronym{iid}{IID}{Independent and Identically Distributed models}
\newacronym{mle}{MLE}{Maximum Likelihood Estimation}
\newacronym{nll}{NLL}{Negative Log-Likelihood}
\newacronym{rcbd}{RCBD}{Randomised Complete Block Design}
\newacronym{ssmu}{SSMU}{Site-specific Management Units}

\usepackage{lineno}
\begin{document}

\begin{frontmatter}



\title{\Large Automatic Quality Control for Agricultural Field Trials\\
\large Detection of Nonstationarity in Grid-indexed Data}


\author{Karen Wolf\corref{cor1}\fnref{inst1,inst2}} %% Author name
\author{Pierre Fernique\fnref{inst1}} %% Author name
\author{Hans-Peter Piepho\fnref{inst2}} %% Author name

\affiliation[inst1]{organization={Limagrain Europe},
%Department and Organization
            addressline={28 Rte d'Ennezat}, 
            city={Chappes},
            postcode={63720}, 
            country={France}}
  
\affiliation[inst2]{organization={University of Hohenheim},%Department and Organization
            addressline={Schloss Hohenheim 1}, 
            city={Stuttgart},
            postcode={70599}, 
            country={Germany}}
            
\cortext[cor1]{Corresponding author: Karen Wolf, \texttt{karen.wolf@uni-hohenheim.de}}

%% Abstract
\begin{abstract}
A common assumption in the spatial analysis of agricultural field trials is stationarity.
In practice, however, this assumption is often violated due to unaccounted field effects.
For instance, in plant breeding field trials, this can lead to inaccurate estimates of plant performance.
Based on such inaccurate estimates, breeders may be impeded in selecting the best performing plant varieties, slowing breeding progress.
We propose a method to automatically verify the hypothesis of stationarity.
The method is sensitive towards mean as well as variance–covariance nonstationarity.
It is specifically developed for the two-dimensional grid-structure of field trials.
The method relies on the hypothesis that we can detect nonstationarity by partitioning the field into areas, within which stationarity holds.
We applied the method to a large number of simulated datasets and a real-data example.
The method reliably points out which trials exhibit quality issues and gives an indication about the severity of nonstationarity.
This information can significantly reduce the time spent on manual quality control and enhance its overall reliability.
Furthermore, the output of the method can be used to improve the analysis of conducted trials as well as the experimental design of future trials.
\end{abstract}

\begin{graphicalabstract}
\includegraphics[width=\textwidth]{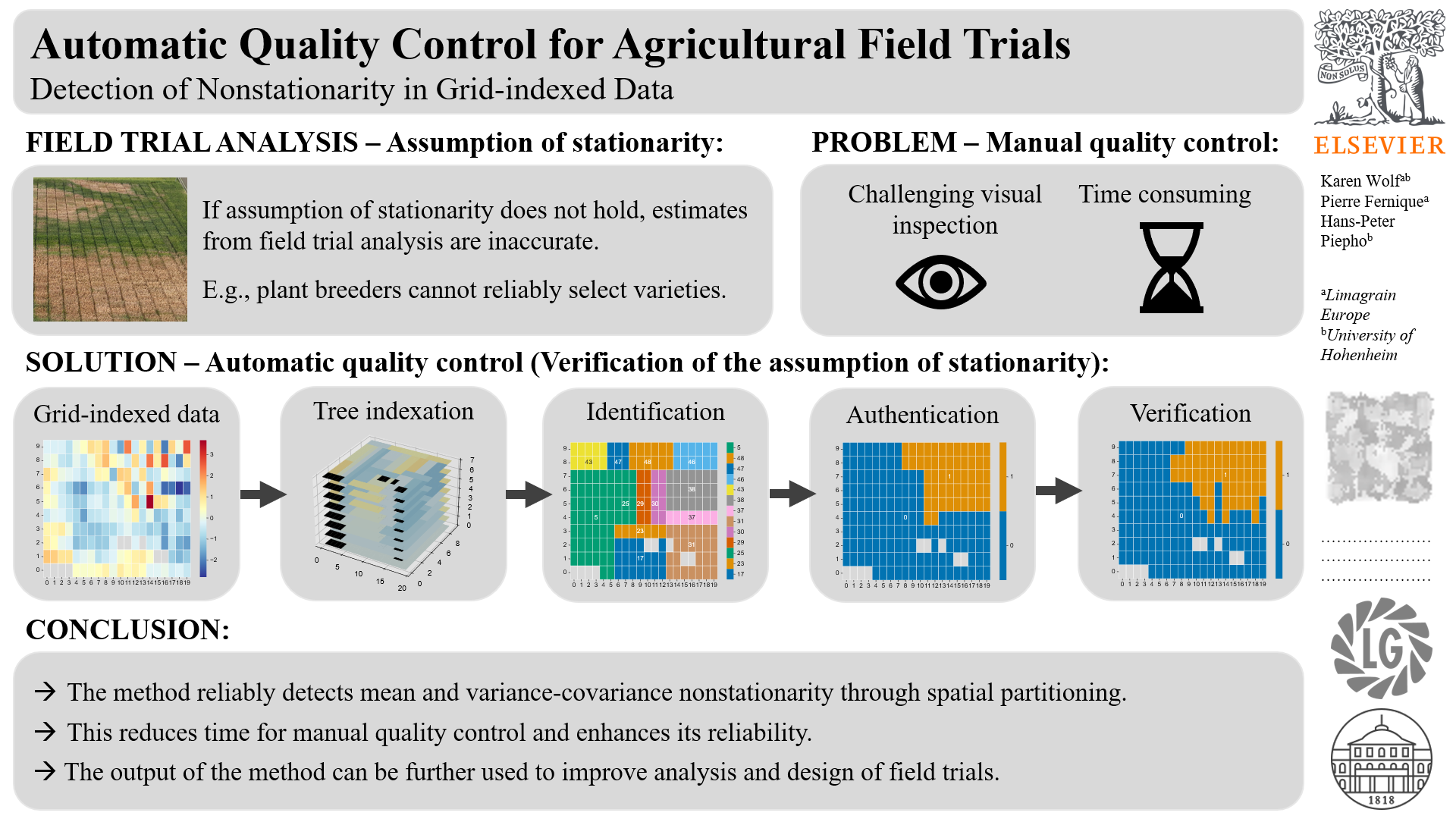}
\end{graphicalabstract}

\begin{highlights}
\item We propose a method for automatic quality control in field trials
\item The method verifies the assumption of mean and variance-covariance stationarity
\item The method indicates the severity of nonstationarity through spatial partitioning
\item The method reduces time for manual quality control and enhances its reliability
\item The method can be used to improve field trial analysis and design
\end{highlights}

%% Keywords

\end{frontmatter}



%% Labels are used to cross-reference an item using \ref command.
% \paragraph{Plant breeding}
%% Use \subsubsection, \paragraph, \subparagraph commands to 
%% start 3rd, 4th and 5th level sections.
%% Refer following link for more details.
%% https://en.wikibooks.org/wiki/LaTeX/Document_Structure#Sectioning_commands

\section{Introduction}
\label{sec:Introduction}
Field trials play a crucial role in providing data to answer agricultural research questions.
We want to highlight the importance of field trials, as well as the importance of their quality, using plant breeding as an example.

\paragraph{Field trials in plant breeding}
Plant breeding aims at producing new plant varieties (i.e., cultivars) with higher performance.
Performance can be defined by agronomic traits such as yield, quality, and the ability to resist to abiotic and biotic factors (e.g., drought, insects) \citep{qaim2020role}.
To achieve this goal, plant breeders must first measure the cultivar performance.
This is done in field trials, where the field is subdivided along two perpendicular axes (i.e., rows and columns) into a regular grid of small, usually rectangular areas (i.e., plots).
Within each plot, a single plant variety is cultivated, while the same variety may be replicated across multiple plots.
For each plot, a trait of interest is measured.
This results in data indexed by rows and columns (i.e., in a two-dimensional grid).

\paragraph{Field trial analysis}
Unfortunately, breeders cannot select for performance directly based on the measured traits.
Field trials are in-vivo experiments and despite a careful selection and management of the field, micro-environmental effects (e.g., variation in soil quality, shadowing of neighbouring plants) are present.
Since each cultivar remains at the same position during the whole cultivation period, the measured values per plot are not only governed by cultivar genetics (i.e., cultivar effect) but also by micro-environmental effects.
All micro-environmental effects are regrouped together into the generic field effect term.

Generally, field trials are analysed in two stages \citep{mohring2009comparison}.
In a first-stage analysis, the aim is to estimate the cultivar effect separately from the undesirable field effect in order to obtain accurate estimations.
The obtained cultivar effect estimates can be used for second-stage analyses.
For example, based on the estimates of several field trials, breeders can compare cultivar performance across multiple locations (i.e., multi-location analysis) \citep{piepho2012stage}.
Moreover, in addition to phenotypic field trial data, genomic data can be included (e.g., Genomic prediction, GWAS) \citep{kumar2024advances}.
However, breeders can only reliably select varieties for high performance if cultivar effect estimation in the first-stage analysis is accurate.

A well-established and straightforward approach for the first-stage analysis of metric traits like yield is to use the framework of \gls{grf} \citep{cressie1993statistics, mao2020adjusting}.
For other traits, as for example ordinal disease scores, this assumption may not hold.
Instead, alternative approaches, such as \gls{glm}, may be considered for the first-stage analysis.
However, in the following, we will focus on the classical analysis method for metric traits.
Other approaches can be equally integrated with the method we propose, as will be discussed in more detail in Section \ref{sec:Discussion}.

A \gls{grf} is denoted as,
\begin{align}
    \boldsymbol{Y} \sim \mathcal{N}(\boldsymbol{X} \boldsymbol{\beta},\,\boldsymbol{\Sigma}),
\label{eq:grf}
\end{align}
where:
\begin{description}
	\item[$\boldsymbol{\beta}$ is the vector of effects:] All effects, along with the cultivar effect that is of highest importance for breeding decisions, can be estimated using the \gls{mle} of $\boldsymbol{\beta}$,
	\begin{align}
		\hat{\boldsymbol{\beta}}_{\text{MLE}} = (\boldsymbol{X}^T \boldsymbol{\Sigma}^{-1} \boldsymbol{X})^{-1} \boldsymbol{X}^T \boldsymbol{\Sigma}^{-1} \boldsymbol{y},
		\label{eq:mle_beta}
	\end{align}
	implied by (\ref{eq:grf}).
	\item[$\boldsymbol{X}$ is the design matrix:] Additionally to the cultivar effect, all a priori known field effects can be taken into account using an experimental design (i.e., specific spatial allocations of randomisations and replications of the cultivars).
	For example, breeders can specify so-called blocks, which are areas in the field within which soil conditions are assumed to be homogeneous \citep{john1995cyclic, bailey2008design}.
	The chosen experimental design is represented by $\boldsymbol{X}$ which encodes for all effects.
	\item[$\boldsymbol{\Sigma}$ is the variance-covariance matrix:] Different types of \gls{grf}s exist, for example \gls{iid} (i.e., $\boldsymbol{\Sigma} = \sigma^2 \mathbb{I}$) or \gls{ar1} (e.g., $\boldsymbol{\Sigma} = \sigma^2 \left(\boldsymbol{\Sigma}_r \otimes \boldsymbol{\Sigma}_c\right)$.
	For more details see \citet{butler2017asreml}.
\end{description}

\paragraph{Nonstationarity}
Note that, apart from the data $\boldsymbol{y}$, the estimation of $\boldsymbol{\beta}$, and thus accuracy of cultivar effect estimation, directly depends on the experimental design $\boldsymbol{X}$ and the variance-covariance matrix $\boldsymbol{\Sigma}$ chosen.
To obtain reliable estimations of $\boldsymbol{\beta}$, we need a parsimonious model (e.g., rank of $\boldsymbol{X}$ is not too large, $\boldsymbol{\Sigma}$ is sparse).
Here, an important hypothesis is stationarity in both mean and variance-covariance.
However, there are several reasons why this assumption may not hold in practice:
\begin{description}
	\item[Not all field effects are a priori known or accounted for in the experimental design.]
	Breeders may not have a priori information about the field, or, need to cope with transient features (e.g., areas of poor drainage showing up only in wet seasons and patches of light soil showing up only in dry ones).
	Then, the chosen experimental design (i.e., $\boldsymbol{X}$) may be inappropriate.
	For example, the blocking structure may not fit the actual field effect pattern \citep{pearce1995some}.
	Blocking then cannot ensure \textit{mean stationarity}.
	\item[Not all field effects can be controlled for by experimental designs.]
	Since the developments of \citet{fisher1935design} in experimental designs in plant breeding, field trials became larger (i.e., contain a high number of varieties to test).
	In this case, 'classical' experimental designs with complete blocks can hardly control for all field effects (e.g., complete blocks are too large to ensure homogeneous field conditions within each block).
	Therefore, alternative designs with incomplete blocks have been developed, such as Lattice designs \citep{yates1936new} and $\alpha$-designs \citep{patterson1978block}.
	For trials without replication of cultivars, Rep-Check designs have been developed \citep{kempton1984design}.
		Still, field correlation within the plots can vary across large distances, making it difficult to assume \textit{variance-covariance stationarity} for the analysis.
		Then, the chosen variance-covariance matrix (i.e., $\boldsymbol{\Sigma}$) may be inappropriate.
\end{description}
In the following, we define nonstationarity as any deviation from the assumption of mean and variance-covariance stationarity.
We assume that in the case of nonstationarity, the grid can be partitioned into disjoint parts, where within each part, the random vector follows a distinct stationary \gls{grf}.
We will refer to such parts as \textit{patches} in the following.

The partition of the two-dimensional grid into patches can be denoted by
\begin{align*}
	\bigcup_{p=1}^m A_p = \mathcal{A},
\end{align*}
where \(A_1, A_2, \dots, A_m\) are \(m\) disjoint patches and \(\mathcal{A}\) denotes the whole grid.
Accordingly, we define \(\boldsymbol{\epsilon}_{A_1}, \boldsymbol{\epsilon}_{A_2}, \dots, \boldsymbol{\epsilon}_{A_m}\) as the subvectors of the vector of residuals (i.e., $\boldsymbol{\epsilon}_{A_p} \in \boldsymbol{\epsilon}, \text{ for } p = 1, \dots, m$), each corresponding to one of the \(m\) disjoint residual patches (resp. \(\boldsymbol{y}_{A_1}, \boldsymbol{y}_{A_2}, \dots, \boldsymbol{y}_{A_m}\)).

We will refer to the parts $\boldsymbol{\epsilon}_{A_p}$ as \textit{residual patches} in the following.
We assume the residuals of different residual patches to be independent and to follow distinct \gls{grf}s, i.e., 
 \begin{align*}
 	\forall\ p \in \{1, \dots, m\} \quad \boldsymbol{\epsilon}_{A_p} \independent \boldsymbol{\epsilon}_{A_q} \text{ if } p \neq q,
 \end{align*}
\begin{align}
 	\forall\ p \in \{1, \dots, m\} \quad
 	\boldsymbol{\epsilon}_{A_p} \sim \mathcal{N}(\boldsymbol{\mu}_{p}, \boldsymbol{\Sigma}_{p}).
 \label{eq:grf_epsilon}
 \end{align}
 
If instead of (\ref{eq:grf_epsilon}) constant expectation and a 'simplistic' covariance structure is assumed (i.e., $\boldsymbol{\epsilon} \sim \mathcal{N}(0,\,\boldsymbol{\Sigma})$),
this will lead to unreliable inferences and the estimation of $\hat{\boldsymbol{\beta}}_{\text{MLE}}$ in (\ref{eq:mle_beta}) will be inaccurate.
 		
\section{Related Work}
\label{sec:Related_Work}
The violation of the assumption of stationarity is a common problem in the analysis of spatial data and beyond.
For example, in time series, stationarity is often assumed but may not hold due to seasonal cycles \citep{wang2024decomposition}.
In spatial datasets, nonstationarity may occur if distances between data points are large, which is often the case in applications such as ecology or geology \citep{osborne2002should}.
Similarly, in field trials, violations of mean as well as variance-covariance stationarity are a known issue \citep{mercer1911experimental}.

There exist various ways to model nonstationary data and many have been applied to field trial datasets \citep{mercer1911experimental, rodriguez2018corr}.
While mean nonstationarity can be tackled by rather simple techniques such as median polishing \citep{cressie1993statistics},
more complex techniques are needed to account for variance-covariance nonstationarity \citep{dreesman2001non}.
They mostly rely on the estimation of spatially-varying parameters, initially termed \gls{gwr} by \citet{brunsdon1996geographically}.
Such modelling techniques can ensure accurate estimations even in the case of nonstationarity.
However, before applying nonstationary modeling techniques, we specifically aim to verify if the assumption of stationarity is actually violated and to which extent (i.e., quality control).
Ideally, the quality control methods are developed in such a way that their output directly supports nonstationary modeling techniques.

\paragraph{Quality control through spatial partitioning}
A possibility for quality control is stationary decomposition.
The pioneering idea of decomposing the state space into stationary components, that can be accounted for in the analysis, was introduced in the frame of time series by \citet{priestley1965evolutionary}.
It has been further developed in its original application \citep{wang2024decomposition} and also adapted to spatial data analysis \citep{dahlhaus2000likelihood}.

Especially for very large spatial datasets, spatial partitioning is advisable \citep{osborne2002should}.
In specific applications, it may not only be an alternative to direct analysis, but a necessity.
For example in geology, changes of the covariance structure may be equally sharp as changes in rock strata across the two-dimensional grid.
In this case, conventional nonstationary analysis techniques, which assume that the covariance between points is a smooth function of distance, are not applicable \citep{kim2005analyzing}.

Although the aim of the spatial partitioning methods developed thus far is very similar to ours, they have mainly been applied to irregularly spaced realisations of data \citep{tzeng2024assessing}.
Therefore, these methods are not directly applicable to agricultural field trials, where data is distributed across a regular grid.

\paragraph{Spatial partitioning in agricultural field trials}
Agricultural research, too, has already recognised and discussed the necessity of spatial partitioning.
For example, \citet{corwin2010delineating} use the term \gls{ssmu} to describe the partition of the field that accounts for spatial variation.
As the term \gls{ssmu} suggests, the aim is to detect spatial variability of factors that influence cultivar performance (e.g., soil physical and chemical properties) to then homogenise the field through location specific management (e.g., fertilisers).
To detect \gls{ssmu}, a sample of metric measures of those factors that are found to influence cultivar performance, are discretised into ordered groups and plotted on the two-dimensional grid.
There are similar approaches which focus on cultivar performance indicators directly.
For example \citet{rakshit2020novel} visualise stationary areas in so-called contour plots for yield in on-farm experiments.
Specifically for field trials, \citet{lacasa2023bayesian} explores nonstationarity, which she defines as any deviation from a stationary \gls{iid} process.

The above mentioned methods for spatial partitioning rely on manual discretisation and focus on mean differences ignoring spatial covariance.
To verify both, mean as well as variance-covariance stationarity, visual inspection is done in practice.
Plant breeders plot the data (i.e., residuals) in heatmaps and manually search for spatial patterns.
Unfortunately, this method is laborious and not all spatial patterns are detectable through visual inspection.
Therefore, it is an unsatisfactory solution, especially for the quality control of large number of trials per season.

\paragraph{Other spatial partitioning techniques for grid-indexed data}
An alternative to manual discretisation is the spatial partitioning method proposed by  \citet{guinness2015likelihood}.
This approach was developed for regular lattice and is based on the Ising model.
The data is interpolated to a grid and can thus be directly transferred to the two-dimensional grid structure of field trial data.
Although the method is well suited for in-depth studies of specific datasets, it may not be easily adaptable for quality control of a large number of datasets, as it requires some fine-tuning.
Furthermore, the method is specifically designed for detecting distinct stationary autocorrelated \gls{gp}.
As a result, it is less suitable for processes without autocorrelation (i.e., \gls{iid}) or for non-Gaussian data types, such as ordinal variables (e.g., disease scoring field trials).

\paragraph{Spatial partitioning using tree graphs}
An alternative approach that more directly accounts for the structure of two-dimensional grids and offers greater flexibility for different types of processes is the use of tree graphs.
\citet{lacasa2023bayesian} uses tree graphs for the assignment of individual plots to blocks with stationary mean (i.e., \gls{iid}).
Thus, she uses the tree only as a classification algorithm for individual plots, ignoring the hierarchical neighbourhood structure.
In contrast, we want to spatially partition the field into larger residual patches containing multiple plots (i.e., compression) of different types of \gls{grf}s (i.e., \gls{iid}, \gls{ar1}).

Tree graphs have been widely applied in image compression for a relatively long time, for example by \citet{hunter1979operations}.
Although Hunter and Steiglitz use in their paper "[...] the word 'picture,' [they] mean any two-dimensional array of information [...].
Likewise, the color of a point may in fact be any information to be associated with a point in a two-dimensional grid." 
Thus, tree indexation cannot only be used to  more compactly represent the information of pictures, but also
to spatially partition field trials.

\citet{gramacy2008bayesian} found tree graphs to be a simple but effective technique to tackle nonstationarity.
In their work, they combine standard \gls{gp} and tree partitioning within a Bayesian framework.
They focus on grid-indexed data, more specifically, on the output of a computer simulation (i.e., NASA rocket booster simulation), but emphasise that the method can be extended to other applications as well.
Furthermore, they point out that, in the frame of \gls{gp}, tree graphs are particularly well-suited for problems with a smaller number of distinct parts.
While \citet{gramacy2008bayesian} use tree indexation to obtain a nonstationary, semiparametric model for analysis, we want to specifically focus on the estimation of partitions in the framework of parametric models.
More precisely, we use tree indexation as a basis for quality control methods in field trials to verify if the assumption of stationarity has been violated (i.e., spatial partitioning into residual patches) and to what extent (i.e., number and size of residual patches).

\section{Proposed Method}
\label{sec:Proposed_Method}
\begin{figure}[H]
\centering
\begin{minipage}{0.45\textwidth}
    % Row 1
    \begin{subfigure}{\textwidth}
        \centering
        \includegraphics[width=0.35\textwidth]{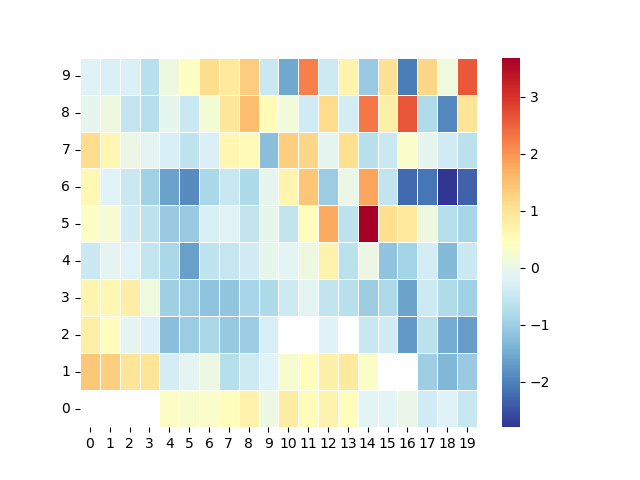}
        \resizebox{0.6\textwidth}{!}{\hspace{0.3cm}        		
        								\begin{tikzpicture}[scale=.38,every node/.style={minimum size=1cm},on grid]

    % Depth 4 (Visible)
    \begin{scope}[
        yshift=0,every node/.append style={
        yslant=0.5,xslant=-1},yslant=0.5,xslant=-1
    ]
        \fill[white,fill opacity=.9] (0,0) rectangle (4,4);
        \draw[black,very thick] (0,0) rectangle (4,4);
        \draw[step=10mm, black] (0,0) grid (4,4);
    \end{scope}

    % Depth 3 (Invisible)
    \begin{scope}[
        yshift=90,every node/.append style={
        yslant=0.5,xslant=-1},yslant=0.5,xslant=-1
    ]
        \fill[white,fill opacity=0] (0,0) rectangle (4,4);
        \draw[draw=none] (0,0) rectangle (4,4);
        \draw[draw=none] (0,1) rectangle (4,2);
        \draw[draw=none] (0,3) rectangle (4,2);
    \end{scope}

    % Depth 2 (Invisible)
    \begin{scope}[
        yshift=170,every node/.append style={
        yslant=0.5,xslant=-1},yslant=0.5,xslant=-1
    ]
        \fill[white,fill opacity=0] (0,0) rectangle (4,4);
        \draw[draw=none] (0,0) rectangle (4,4);
    \end{scope}

    % Depth 1 (Invisible)
    \begin{scope}[
        yshift=240,every node/.append style={
        yslant=0.5,xslant=-1},yslant=0.5,xslant=-1
    ]
        \fill[white,fill opacity=0] (0,0) rectangle (4,4);
        \draw[draw=none] (0,0) rectangle (4,4);
    \end{scope}

    % Depth 0 (Invisible)
    \begin{scope}[
        yshift=310,every node/.append style={
        yslant=0.5,xslant=-1},yslant=0.5,xslant=-1
    ]
        \fill[white,fill opacity=0] (0,0) rectangle (4,4);
        \draw[draw=none] (0,0) rectangle (4,4);
    \end{scope}

\end{tikzpicture}
            							\hspace{7.5cm}}
        \caption[Subfigure 1a]{\label{subfig:1a}}
        \tikz[remember picture] \node (A) {};
    \end{subfigure}
    % Arrow
    \tikz[remember picture, overlay]
        \draw[->, thick]
            ($(A.east) + (-2mm,0)$) -- ++(0,-0.25cm) node[midway, above] {};
    
    % Row 2
    \begin{subfigure}{\textwidth}
        \centering
        \includegraphics[width=0.35\textwidth]{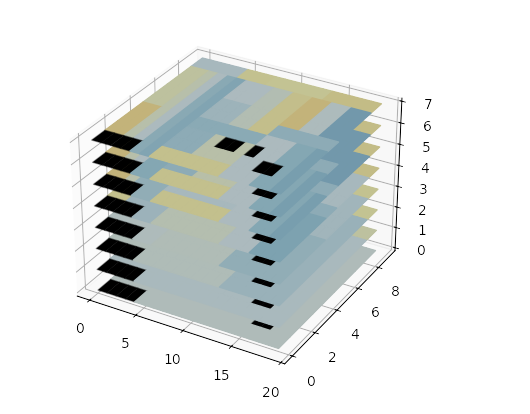}
        \resizebox{0.6\textwidth}{!}{\hspace{0.3cm}
            							\begin{tikzpicture}[scale=.38,every node/.style={minimum size=1cm},on grid]

	% depth 4
    \begin{scope}[
    		yshift=0,every node/.append style={
    	    yslant=0.5,xslant=-1},yslant=0.5,xslant=-1
    	             ]
        \fill[white,fill opacity=.9] (0,0) rectangle (4,4);
        \draw[black,very thick] (0,0) rectangle (4,4);
        \draw[step=10mm, black] (0,0) grid (4,4);
    \end{scope}
    
	% depth 3
    \begin{scope}[
    		yshift=90,every node/.append style={
    		yslant=0.5,xslant=-1},yslant=0.5,xslant=-1
    	             ]
    		\fill[white,fill opacity=.9] (0,0) rectangle (4,4);
    		\draw[step=20mm, black] (0,0) grid (4,4);
    		\draw[black] (0,1) rectangle (4,2);
    		\draw[black] (0,3) rectangle (4,2);
    		\draw[black,very thick] (0,0) rectangle (4,4);
    \end{scope}
    
	% depth 2
    \begin{scope}[
    		yshift=170,every node/.append style={
    	    yslant=0.5,xslant=-1},yslant=0.5,xslant=-1
    	  ]
        \fill[white,fill opacity=0.6] (0,0) rectangle (4,4);
        \draw[step=20mm, black] (0,0) grid (4,4);
        \draw[black,very thick] (0,0) rectangle (4,4);
    \end{scope}
    
	% depth 1
    \begin{scope}[
    		yshift=240,every node/.append style={
    	    yslant=0.5,xslant=-1},yslant=0.5,xslant=-1
    	  ]
        \fill[white,fill opacity=0.6] (0,0) rectangle (4,4);
        \draw[step=40mm, black] (0,0) grid (4,4);
        \draw[black] (0,2) rectangle (4,2);
        \draw[black,very thick] (0,0) rectangle (4,4);
    \end{scope}
    
   	% depth 0     
    \begin{scope}[
   	 	yshift=310,every node/.append style={
    	    yslant=0.5,xslant=-1},yslant=0.5,xslant=-1
    	  ]
        \fill[white,fill opacity=0.6] (0,0) rectangle (4,4);
        \draw[step=40mm, black] (0,0) grid (4,4);
        \draw[black,very thick] (0,0) rectangle (4,4);
    \end{scope}    

\end{tikzpicture}
            							\hspace{1cm}
            							\begin{tikzpicture}[scale=0.3, transform shape,
					roundnode/.style={circle, draw=black!75, fill=black!5, very thick, minimum size=1cm},]

    % depth 0
    \node[roundnode] (node0)                            				{0};

    % depth 1
    \node[roundnode] (node1) [below left=2cm and 4.6cm of node0]		{1};
    \node[roundnode] (node2) [below right=2cm and 4.6cm of node0]		{2};
    \draw[->] 	(node0.south) -- (node1.north);
    \draw[->] 	(node0.south) -- (node2.north);

    % depth 2
    \node[roundnode] (node3) [below left=3cm and 1.9cm of node1]		{3};
    \node[roundnode] (node4) [below right=3cm and 1.9cm of node1]		{4};
    \draw[->]	(node1.south) -- (node3.north);
    \draw[->]	(node1.south) -- (node4.north);
        
    \node[roundnode] (node5) [below left=3cm and 1.9cm of node2]		{5};
    \node[roundnode] (node6) [below right=3cm and 1.9cm of node2]		{6};
    \draw[->]			(node2.south) -- (node5.north);
    \draw[->]			(node2.south) -- (node6.north);
    
    % depth 3
    \node[roundnode] (node7) [below left=4cm and 0.6cm of node3]		{7};
    \node[roundnode] (node8) [below right=4cm and 0.6cm of node3]		{8};
    \draw[->]			(node3.south) -- (node7.north);
    \draw[->]			(node3.south) -- (node8.north);
        
    \node[roundnode] (node9) [below left=4cm and 0.6cm of node4]		{9};
    \node[roundnode] (node10) [below right=4cm and 0.6cm of node4]	{10};
    \draw[->]	(node4.south) -- (node9.north);
    \draw[->]	(node4.south) -- (node10.north);
        
    \node[roundnode] (node11) [below left=4cm and 0.6cm of node5]		{11};
    \node[roundnode] (node12) [below right=4cm and 0.6cm of node5]	{12};
    \draw[->]			(node5.south) -- (node11.north);
    \draw[->]			(node5.south) -- (node12.north);
        
    \node[roundnode] (node13) [below left=4cm and 0.6cm of node6]		{13};
    \node[roundnode] (node14) [below right=4cm and 0.6cm of node6]	{14};
    \draw[->]	(node6.south) -- (node13.north);
    \draw[->]	(node6.south) -- (node14.north);
    
    % depth 4
    \node[roundnode] (node15) [below left=5cm and 0cm of node7]		{15};
    \node[roundnode] (node16) [below right=5cm and 0cm of node7]		{16};
    \draw[->]			(node7.south) -- (node15.north);
    \draw[->]			(node7.south) -- (node16.north);
        
    \node[roundnode] (node17) [below left=5cm and 0cm of node8]		{17};
    \node[roundnode] (node18) [below right=5cm and 0cm of node8]		{18};
    \draw[->]			(node8.south) -- (node17.north);
    \draw[->]			(node8.south) -- (node18.north);
        
    \node[roundnode] (node19) [below left=5cm and 0cm of node9]		{19};
    \node[roundnode] (node20) [below right=5cm and 0cm of node9]		{20};
    \draw[->]			(node9.south) -- (node19.north);
    \draw[->]			(node9.south) -- (node20.north);
        
    \node[roundnode] (node21) [below left=5cm and 0cm of node10]		{21};
    \node[roundnode] (node22) [below right=5cm and 0cm of node10]		{22};
    \draw[->]	(node10.south) -- (node21.north);
    \draw[->]	(node10.south) -- (node22.north);
        
    \node[roundnode] (node23) [below left=5cm and 0cm of node11]		{23};
    \node[roundnode] (node24) [below right=5cm and 0cm of node11]		{24};
    \draw[->]			(node11.south) -- (node23.north);
    \draw[->]			(node11.south) -- (node24.north);
        
    \node[roundnode] (node25) [below left=5cm and 0cm of node12]		{25};
    \node[roundnode] (node26) [below right=5cm and 0cm of node12]		{26};
    \draw[->]			(node12.south) -- (node25.north);
    \draw[->]			(node12.south) -- (node26.north);
        
    \node[roundnode] (node27) [below left=5cm and 0cm of node13]		{27};
    \node[roundnode] (node28) [below right=5cm and 0cm of node13]		{28};
    \draw[->]			(node13.south) -- (node27.north);
    \draw[->]			(node13.south) -- (node28.north);
        
    \node[roundnode] (node29) [below left=5cm and 0cm of node14]		{29};
    \node[roundnode] (node30) [below right=5cm and 0cm of node14]		{30};
    \draw[->]			(node14.south) -- (node29.north);
    \draw[->]			(node14.south) -- (node30.north);

\end{tikzpicture}}
        \caption[Subfigure 1b]{\label{subfig:1b}}
        \tikz[remember picture] \node (B) {};
    \end{subfigure}
    % Arrow
    \tikz[remember picture, overlay]
        \draw[->, thick]
            ($(B.east) + (-2mm,0)$) -- ++(0,-0.25cm) node[midway, above] {};

    % Row 3
    \begin{subfigure}{\textwidth}
        \centering
        \includegraphics[width=0.35\textwidth]{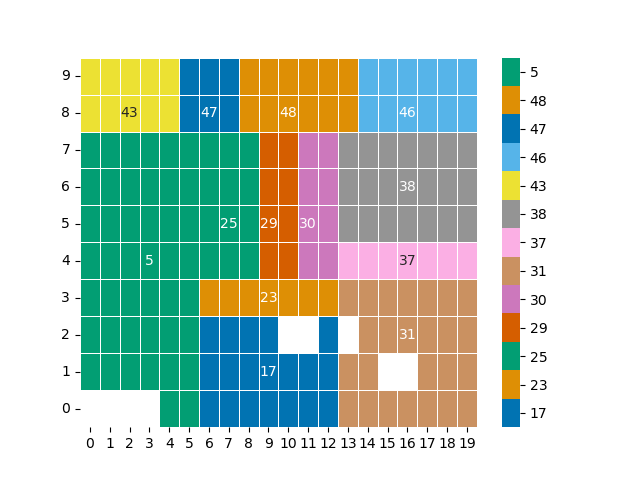}
        \resizebox{0.6\textwidth}{!}{\hspace{0.3cm}
            							\begin{tikzpicture}[scale=.38,every node/.style={minimum size=1cm},on grid]

	% depth 4
    \begin{scope}[
           yshift=0,every node/.append style={
           yslant=0.5,xslant=-1},yslant=0.5,xslant=-1
        ]
        	\fill[white,fill opacity=.9] (0,0) rectangle (4,4);
        \draw[step=10mm, black, opacity=.9] (0,0) grid (4,4);
        \fill[Lavender] (2,1) rectangle (3,2);
        \draw[black,very thick] (2,1) rectangle (3,2);
        \fill[ForestGreen] (3,1) rectangle (4,2);
        \draw[black,very thick] (3,1) rectangle (4,2);
        \fill[SeaGreen] (0,3) rectangle (1,4);
        \draw[black,very thick] (0,3) rectangle (1,4);
        \fill[RubineRed] (1,3) rectangle (2,4);
        \draw[black,very thick] (1,3) rectangle (2,4);
    \end{scope}
    
	% depth 3
    \begin{scope}[
    		yshift=90,every node/.append style={
    		yslant=0.5,xslant=-1},yslant=0.5,xslant=-1
    	             ]
    	    \draw[black,very thick] (0,0) rectangle (4,4);
        \fill[YellowGreen] (0,0) rectangle (1,1);
        \fill[YellowGreen] (1,0) rectangle (2,1);
    	    \draw[black,very thick] (0,0) rectangle (2,1);
        \fill[SkyBlue] (2,0) rectangle (3,1);
        \fill[SkyBlue] (3,0) rectangle (4,1);
    	    \draw[black,very thick] (2,0) rectangle (4,1);
        \fill[Cerulean] (0,1) rectangle (1,2);
        \fill[Cerulean] (1,1) rectangle (2,2);
    	    \draw[black,very thick] (0,1) rectangle (2,2);
        \fill[Aquamarine] (0,2) rectangle (1,3);
        \fill[Aquamarine] (1,2) rectangle (2,3);
        \draw[black,very thick] (0,2) rectangle (2,3);
        \fill[Salmon] (2,2) rectangle (3,3);
        \fill[Salmon] (3,2) rectangle (4,3);
        \draw[black,very thick] (2,2) rectangle (4,3);
        \fill[Red] (2,3) rectangle (3,4);
        \fill[Red] (3,3) rectangle (4,4);
        \draw[black,very thick] (2,3) rectangle (4,4);
    \end{scope}
    
	% depth 2
    \begin{scope}[
    		yshift=170,every node/.append style={
    	    yslant=0.5,xslant=-1},yslant=0.5,xslant=-1
    	  ]
        \fill[white,fill opacity=0.6] (0,0) rectangle (4,4);
        \draw[step=20mm, black] (0,0) grid (4,4);
        \draw[black,very thick] (0,0) rectangle (4,4);
    \end{scope}

	% depth 1
    \begin{scope}[
    		yshift=240,every node/.append style={
    	    yslant=0.5,xslant=-1},yslant=0.5,xslant=-1
    	  ]
        \fill[white,fill opacity=0.6] (0,0) rectangle (4,4);
        \draw[step=40mm, black] (0,0) grid (4,4);
        \draw[black] (0,2) rectangle (4,2);
        \draw[black,very thick] (0,0) rectangle (4,4);
    \end{scope}

   	% depth 0     
    \begin{scope}[
   	 	yshift=310,every node/.append style={
    	    yslant=0.5,xslant=-1},yslant=0.5,xslant=-1
    	  ]
        \fill[white,fill opacity=0.6] (0,0) rectangle (4,4);
        \draw[step=40mm, black] (0,0) grid (4,4);
        \draw[black,very thick] (0,0) rectangle (4,4);
    \end{scope}    

\end{tikzpicture}
            							\hspace{1cm}
            							\begin{tikzpicture}[scale=0.3, transform shape,
					roundnode/.style={circle, draw=black!75, fill=black!5, very thick, minimum size=1cm},
					Lroundnode/.style={circle, draw=Lavender!75, fill=black!5, very thick, minimum size=1cm},
					Aroundnode/.style={circle, draw=Aquamarine!75, fill=black!5, very thick, minimum size=1cm},
					Croundnode/.style={circle, draw=Cerulean!75, fill=black!5, very thick, minimum size=1cm},
					YGroundnode/.style={circle, draw=YellowGreen!75, fill=black!5, very thick, minimum size=1cm},
					FGroundnode/.style={circle, draw=ForestGreen!75, fill=black!5, very thick, minimum size=1cm},
					Rroundnode/.style={circle, draw=Red!75, fill=black!5, very thick, minimum size=1cm},
					RRroundnode/.style={circle, draw=RubineRed!75, fill=black!5, very thick, minimum size=1cm},
					Sroundnode/.style={circle, draw=Salmon!75, fill=black!5, very thick, minimum size=1cm},
					Droundnode/.style={circle, draw=SeaGreen!75, fill=black!5, very thick, minimum size=1cm},
					SBroundnode/.style={circle, draw=SkyBlue!75, fill=black!5, very thick, minimum size=1cm},]

    % depth 0
        \node[roundnode] (node0)                            				{0};
    
    % depth 1
        \node[roundnode] (node1) [below left=2cm and 4.6cm of node0]		{1};
        \node[roundnode] (node2) [below right=2cm and 4.6cm of node0]		{2};
        \draw[->] 	(node0.south) -- (node1.north);
        \draw[->] 	(node0.south) -- (node2.north);
    
    % depth 2
        \node[roundnode] (node3) [below left=3cm and 1.9cm of node1]		{3};
        \node[roundnode] (node4) [below right=3cm and 1.9cm of node1]		{4};
        \draw[->]	(node1.south) -- (node3.north);
        \draw[->]	(node1.south) -- (node4.north);
        
        \node[roundnode] (node5) [below left=3cm and 1.9cm of node2]		{5};
        \node[roundnode] (node6) [below right=3cm and 1.9cm of node2]		{6};
        \draw[->]			(node2.south) -- (node5.north);
        \draw[->]			(node2.south) -- (node6.north);
    
    % depth 3
        \node[roundnode] (node7) [below left=4cm and 0.6cm of node3]		{7};
        \node[Rroundnode] (node8) [below right=4cm and 0.6cm of node3]	{8};
        \draw[->]			(node3.south) -- (node7.north);
        \draw[->]			(node3.south) -- (node8.north);
        
        \node[Aroundnode] (node9) [below left=4cm and 0.6cm of node4]		{9};
        \node[Sroundnode] (node10) [below right=4cm and 0.6cm of node4]	{10};
        \draw[->]	(node4.south) -- (node9.north);
        \draw[->]	(node4.south) -- (node10.north);
        
        \node[Croundnode] (node11) [below left=4cm and 0.6cm of node5]	{11};
        \node[roundnode] (node12) [below right=4cm and 0.6cm of node5]	{12};
        \draw[->]			(node5.south) -- (node11.north);
        \draw[->]			(node5.south) -- (node12.north);

	\node[YGroundnode] (node13) [below left=4cm and 0.6cm of node6]	{[13]};
	\node[SBroundnode] (node14) [below right=4cm and 0.6cm of node6]	{[14]};
	\draw[->]	(node6.south) -- (node13.north);
	\draw[->]	(node6.south) -- (node14.north);

    % depth 4
        \node[Droundnode] (node15) [below left=5cm and 0cm of node7]		{15};
        \node[RRroundnode] (node16) [below right=5cm and 0cm of node7]	{16};
        \draw[->]			(node7.south) -- (node15.north);
        \draw[->]			(node7.south) -- (node16.north);
        
        \node[Lroundnode] (node25) [below left=5cm and 0cm of node12]		{25};
        \node[FGroundnode] (node26) [below right=5cm and 0cm of node12]	{26};
        \draw[->]			(node12.south) -- (node25.north);
        \draw[->]			(node12.south) -- (node26.north);

\end{tikzpicture}}
        \caption[Subfigure 1c]{\label{subfig:1c}}
    \end{subfigure}
\end{minipage}
\hfill
\begin{minipage}{0.45\textwidth}
    % Arrow
    \tikz[remember picture, overlay]
    \draw[->, thick]
    		(2.5cm,0cm) -- ++(0cm, -0.25cm);
    	\vspace{0.3cm}
    		
    % Row 4
    \begin{subfigure}{\textwidth}
        \centering
        \includegraphics[width=0.35\textwidth]{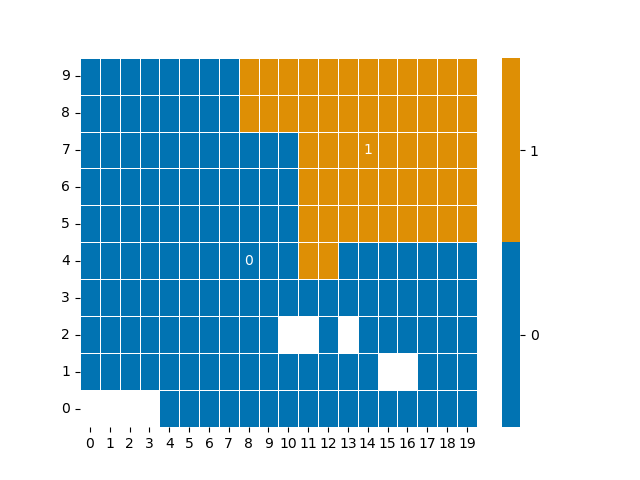}
        \resizebox{0.6\textwidth}{!}{\hspace{0.3cm}
            							\begin{tikzpicture}[scale=.38,every node/.style={minimum size=1cm},on grid]
    
	% depth 4
        \begin{scope}[
            yshift=0,every node/.append style={
            yslant=0.5,xslant=-1},yslant=0.5,xslant=-1
        ]
            \fill[white,fill opacity=.9] (0,0) rectangle (4,4);
            \draw[step=10mm, black] (0,0) grid (4,4);
            \fill[Blue] (0,0) rectangle (1,1);
            \fill[Blue] (1,0) rectangle (2,1);
            \fill[Blue] (2,0) rectangle (3,1);
            \fill[Blue] (3,0) rectangle (4,1);
            \fill[Blue] (0,1) rectangle (1,2);
            \fill[Blue] (1,1) rectangle (2,2);
            \fill[Orange] (2,1) rectangle (3,2);
            \fill[Blue] (3,1) rectangle (4,2);
            \fill[Blue] (0,2) rectangle (1,3);
            \fill[Blue] (1,2) rectangle (2,3);
            \fill[Orange] (2,2) rectangle (3,3);
            \fill[Orange] (3,2) rectangle (4,3);
            \fill[Blue] (0,3) rectangle (1,4);
            \fill[Orange] (1,3) rectangle (2,4);
            \fill[Orange] (2,3) rectangle (3,4);
            \fill[Orange] (3,3) rectangle (4,4);
            \draw[very thick, black] (0,0) -- (4,0) -- (4,2) -- (3,2) -- (3,1) -- (2,1) -- (2,3) -- (1,3) -- (1,4) -- (0,4) -- cycle;
            \draw[black,very thick] (0,0) rectangle (4,4);
        \end{scope}
    
	% depth 3
    \begin{scope}[
    		yshift=90,every node/.append style={
    		yslant=0.5,xslant=-1},yslant=0.5,xslant=-1
    	             ]
    		\fill[white,fill opacity=.9] (0,0) rectangle (4,4);
    		\draw[step=20mm, black] (0,0) grid (4,4);
    		\draw[black] (0,1) rectangle (4,2);
    		\draw[black] (0,3) rectangle (4,2);
    		\draw[black,very thick] (0,0) rectangle (4,4);
    \end{scope}
    
	% depth 2
    \begin{scope}[
    		yshift=170,every node/.append style={
    	    yslant=0.5,xslant=-1},yslant=0.5,xslant=-1
    	  ]
        \fill[white,fill opacity=0.6] (0,0) rectangle (4,4);
        \draw[step=20mm, black] (0,0) grid (4,4);
        \draw[black,very thick] (0,0) rectangle (4,4);
    \end{scope}
    
	% depth 1
    \begin{scope}[
    		yshift=240,every node/.append style={
    	    yslant=0.5,xslant=-1},yslant=0.5,xslant=-1
    	  ]
        \fill[white,fill opacity=0.6] (0,0) rectangle (4,4);
        \draw[step=40mm, black] (0,0) grid (4,4);
        \draw[black] (0,2) rectangle (4,2);
        \draw[black,very thick] (0,0) rectangle (4,4);
    \end{scope}
    
   	% depth 0     
    \begin{scope}[
   	 	yshift=310,every node/.append style={
    	    yslant=0.5,xslant=-1},yslant=0.5,xslant=-1
    	  ]
        \fill[white,fill opacity=0.6] (0,0) rectangle (4,4);
        \draw[step=40mm, black] (0,0) grid (4,4);
        \draw[black,very thick] (0,0) rectangle (4,4);
    \end{scope}    

\end{tikzpicture}
            							\hspace{1cm}
            							\begin{tikzpicture}[scale=0.3, transform shape,
					roundnode/.style={circle, draw=black!75, fill=black!5, very thick, minimum size=1cm},
					Lroundnode/.style={circle, draw=Orange!75, fill=black!5, very thick, minimum size=1cm},
					Aroundnode/.style={circle, draw=Blue!75, fill=black!5, very thick, minimum size=1cm},
					Croundnode/.style={circle, draw=Blue!75, fill=black!5, very thick, minimum size=1cm},
					YGroundnode/.style={circle, draw=Blue!75, fill=black!5, very thick, minimum size=1cm},
					FGroundnode/.style={circle, draw=Blue!75, fill=black!5, very thick, minimum size=1cm},
					Rroundnode/.style={circle, draw=Orange!75, fill=black!5, very thick, minimum size=1cm},
					RRroundnode/.style={circle, draw=Orange!75, fill=black!5, very thick, minimum size=1cm},
					Sroundnode/.style={circle, draw=Orange!75, fill=black!5, very thick, minimum size=1cm},
					Droundnode/.style={circle, draw=Blue!75, fill=black!5, very thick, minimum size=1cm},]

    % depth 0
    \node[roundnode] (node0)                            				{0};

    % depth 1
    \node[roundnode] (node1) [below left=2cm and 4.6cm of node0]		{1};
    \node[roundnode] (node2) [below right=2cm and 4.6cm of node0]		{2};
    \draw[->] 	(node0.south) -- (node1.north);
    \draw[->] 	(node0.south) -- (node2.north);

    % depth 2
    \node[roundnode] (node3) [below left=3cm and 1.9cm of node1]		{3};
    \node[roundnode] (node4) [below right=3cm and 1.9cm of node1]		{4};
    \draw[->]	(node1.south) -- (node3.north);
    \draw[->]	(node1.south) -- (node4.north);
        
    \node[roundnode] (node5) [below left=3cm and 1.9cm of node2]		{5};
    \node[YGroundnode] (node6) [below right=3cm and 1.9cm of node2]	{6};
    \draw[->]			(node2.south) -- (node5.north);
    \draw[->]			(node2.south) -- (node6.north);
    
    % depth 3
    \node[roundnode] (node7) [below left=4cm and 0.6cm of node3]		{7};
    \node[Rroundnode] (node8) [below right=4cm and 0.6cm of node3]	{8};
    \draw[->]			(node3.south) -- (node7.north);
    \draw[->]			(node3.south) -- (node8.north);
        
    \node[Aroundnode] (node9) [below left=4cm and 0.6cm of node4]		{9};
    \node[Sroundnode] (node10) [below right=4cm and 0.6cm of node4]	{10};
    \draw[->]	(node4.south) -- (node9.north);
    \draw[->]	(node4.south) -- (node10.north);
        
    \node[Croundnode] (node11) [below left=4cm and 0.6cm of node5]	{11};
    \node[roundnode] (node12) [below right=4cm and 0.6cm of node5]	{12};
    \draw[->]			(node5.south) -- (node11.north);
    \draw[->]			(node5.south) -- (node12.north);

    % depth 4
    \node[Droundnode] (node15) [below left=5cm and 0cm of node7]		{15};
    \node[RRroundnode] (node16) [below right=5cm and 0cm of node7]	{16};
    \draw[->]			(node7.south) -- (node15.north);
    \draw[->]			(node7.south) -- (node16.north);
        
    \node[Lroundnode] (node25) [below left=5cm and 0cm of node12]		{25};
    \node[FGroundnode] (node26) [below right=5cm and 0cm of node12]	{26};
    \draw[->]			(node12.south) -- (node25.north);
    \draw[->]			(node12.south) -- (node26.north);
        
\end{tikzpicture}}
        \caption[Subfigure 1d]{\label{subfig:1d}}
        \tikz[remember picture] \node (D) {};
    \end{subfigure}
    % Arrow
    \tikz[remember picture, overlay]
        \draw[->, thick]
            ($(D.east) + (-2mm,0)$) -- ++(0,-0.25cm) node[midway, above] {};

    % Row 5
    \begin{subfigure}{\textwidth}
        \centering
        \includegraphics[width=0.35\textwidth]{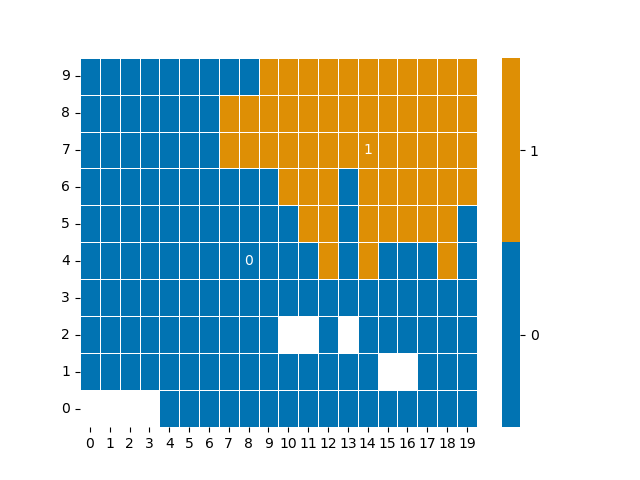}
        \resizebox{0.6\textwidth}{!}{\hspace{0.3cm}
            							\begin{tikzpicture}[scale=.38,every node/.style={minimum size=1cm},on grid]

    % depth 4
    \begin{scope}[
        yshift=0,every node/.append style={
        yslant=0.5,xslant=-1},yslant=0.5,xslant=-1
    ]
        \fill[white,fill opacity=.9] (0,0) rectangle (4,4);
        \draw[step=10mm, black] (0,0) grid (4,4);
        \fill[Blue] (0,0) rectangle (1,1);
        \fill[Blue] (1,0) rectangle (2,1);
        \fill[Blue] (2,0) rectangle (3,1);
        \fill[Blue] (3,0) rectangle (4,1);
        \fill[Blue] (0,1) rectangle (1,2);
        \fill[Blue] (1,1) rectangle (2,2);
        \fill[Orange] (2,1) rectangle (3,2);
        \fill[Blue] (3,1) rectangle (4,2);
        \fill[Blue] (0,2) rectangle (1,3);
        \fill[Orange] (1,2) rectangle (2,3);
        \fill[Orange] (2,2) rectangle (3,3);
        \fill[Orange] (3,2) rectangle (4,3);
        \fill[Blue] (0,3) rectangle (1,4);
        \fill[Blue] (1,3) rectangle (2,4);
        \fill[Orange] (2,3) rectangle (3,4);
        \fill[Orange] (3,3) rectangle (4,4);
        \draw[very thick, black] (0,0) -- (4,0) -- (4,2) -- (3,2) -- (3,1) -- (2,1) -- (2,2) -- (1,2) -- (1,3) -- (2,3) -- (2,4);
        \draw[black,very thick] (0,0) rectangle (4,4);
    \end{scope}

\end{tikzpicture}

% DAS HIER SIND DIE ZEILEN; VON BIS \fill[teal, fill opacity=0.6] (4,0)
% DAS HIER SIND DIE SPALTEN VON BIS  rectangle (0,2);

%\captionof{figure}{Illustration of data segmentation using grid representation of an exemplary trial with four rows and four columns}
%\label{fig:GridRepresentationTree}
            							\hspace{7.5cm}}
        \caption[Subfigure 1e]{\label{subfig:1e}}
        \tikz[remember picture] \node (E) {};
    \end{subfigure}

    % Loop-back Arrow
    \tikz[remember picture, overlay]
        \draw[->, thick]
            (E.east) -- ++(0,-0.3cm) -- ++(3cm,0) -- ++(0,5cm) -- ++(-0.5cm,0) node[midway, above] {};
\end{minipage}
    \caption[Figure 1]
    {\label{fig:1}
    Detection of nonstationarity (i.e., residual patches).
    Data example with spatial patterns (a).
    Partitioning of the data through tree indexation (b).
    Identification of parts (c).
    Merging of parts through authentication (d).
    Rechallenging of the borders of the detected patches through verification (e).}
\end{figure}
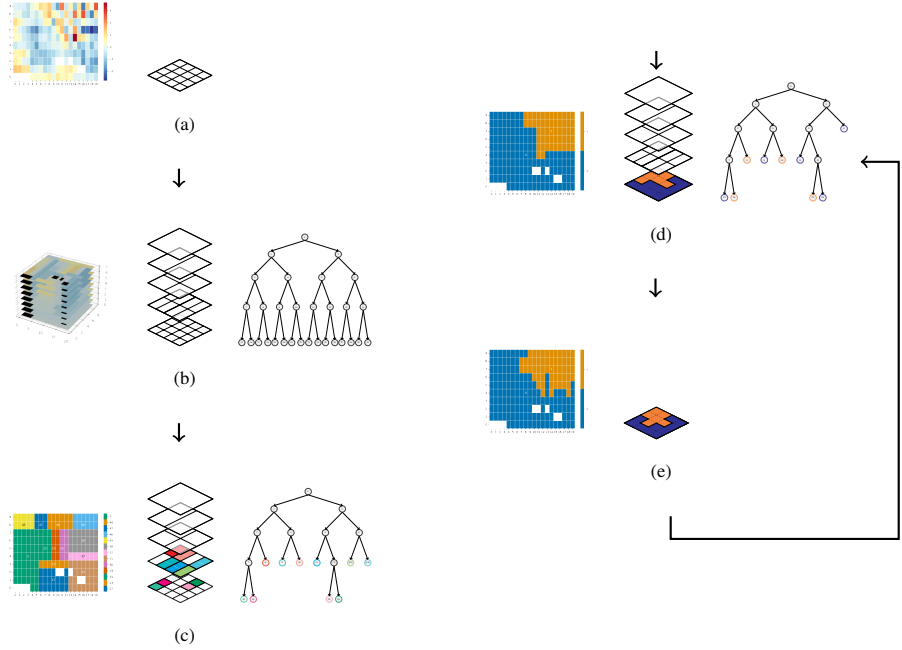

To verify the assumption of stationarity for both mean and variance-covariance (i.e., spatial partitioning into residual patches), we use tree indexation and subsequently apply algorithms for identification, authentication and verification:

\begin{enumerate}
\item We use \textbf{tree indexation} to obtain a hierarchical multi-scale representation of the two-dimensional grid data.
To do this, we iteratively partition the field, where each part of a partition is repartitioned in the next iteration (i.e., nested partition).
The first iteration corresponds to a partition with only one part and thus the least detailed representation of the data (i.e., coarsest scale).
Higher levels of detail (i.e., finer scales) correspond to less compact representations where the field is partitioned into disjoint areas.
At the most detailed representation (i.e., finest scale), each part consists of a single plot (see Figure \ref{subfig:1b}).
\item To \textbf{identify} the best representation of the data among all possibilities of the multi-scale representation, we use a heuristic approach.
We assume that the partition of the field into residual patches lies within the extremes of the coarsest and the finest scale of the multi-scale representation.
To find this partition, we can traverse the tree from the coarsest to the finest scale and determine whether splitting a part into the elements of its nested partition provides a better representation of the data.
Or, conversely, we traverse the tree from the finest to the coarsest scale and determine whether merging the parts of a nested partition provides a better representation of the data (see Figure \ref{subfig:1c}).
\item In doing so, we obtain a representation that aims to strike a balance between representing the data as compactly as possible and as detailed as necessary.
However, the decision of splitting or merging parts of a nested partition is independent of all other parts.
Thus, by construction, we expect to identify too many patches.
We must therefore assess (i.e., \textbf{authenticate}) whether patches can be further merged to obtain the most compact representation of the full data (see Figure \ref{subfig:1d}).
\item The shape of the authenticated patches is highly dependent on the tree indexation (i.e., division of the field parallel to its axes).
Thus, although the number and location of the detected patches might be close to the true patches, their shape may not be realistic.
To \textbf{verify} the true shape of the detected patches, we therefore rechallenge their bordering plots iteratively (see Figure \ref{subfig:1e}).
\end{enumerate}

We repeat authentication and verification until we obtain the number and shape of patches that adequately represent the data within the different parts of the field.
Note that, in a first cycle of authentication and verification, we restrict authentication to solely merge parts that are direct neighbours (i.e., local authentication).
This allows us to find parts of sufficient size.
In a second cycle of authentication and verification, we remove this restriction to obtain the most compact representation of the data (i.e., global authentication).
We always perform a single round of local authentication before each global authentication to avoid the estimation of too many small, disjoint patches.
Each cycle runs until convergence.

A last cycle of global authentication can be applied to control the level of stringency of authentication.
Similarly to the second cycle of authentication, each global authentication is preceded by a single round of local authentication and followed by verification.
The level of stringency of this last authentication can be set by the user.

Nearly all steps of the method involve model comparison or model selection.
For that purpose, we use use penalised scores such as \gls{bic}, \gls{aic}, \gls{nll}, and \gls{bf}.
\gls{nll} is calculated based on cross-validation.
\gls{bf} is calculated based on \gls{nll}.
For each step, a different score may be used.
To control the level of stringency of the last cycle of authentication, we specifically use \gls{bf}.
The level of stringency can be set 
by applying different thresholds such as 'anecdotal' or 'decisive' \citep{krass1995bayes}.

\subsection{Tree indexation}
\label{subsec:Tree_indexation}

\paragraph{Definition of a tree graph}
\label{paragraph:Definition_of_a_tree graph}
In graph theory \citep{lauritzen1996graphical}, a graph $G$ is defined by two sets, vertices $V \subset \mathbb{N}$ and edges $E$, with
\begin{align*}
& \emptyset \subseteq E \subseteq \left \{ (u,v) \in V^2 \mid u \neq v \right \}.
\end{align*}
To apply the concept of tree graphs to the indexation of grid-indexed data, some definitions are needed:
\begin{align*}
  \textbf{child set:} \quad 
  & \forall v \in V, ch(v) = \left \{ u \in V \mid (v,u) \in E \right \},\\
  \textbf{descendant set:} \quad 
  & \forall v \in V, de(v) = \left \{ \cup_{u \in ch(v)}de(u) \right \} \cup ch(v),\\
  \textbf{parent set:} \quad 
  & \forall v \in V, pa(v) = \left \{u \in V \mid (u,v) \in E \right \},\\
  \textbf{sibling set:} \quad 
  & \forall v \in V, si(v) = \left \{ u \in V \mid u \in ch(pa(v)),\ u \neq v \right \},\\
  \textbf{leaf set:} \quad 
  & \forall v \in V, le(v) = \left \{ u \in de(v) \mid ch(u) = \emptyset \right \},\\
  & \forall A \subseteq V, le(A) =  \cup_{v \in A} le(v),
\end{align*}
referring either to a single vertex $v$ or a set of vertices $A$.

We associate a set of coordinates to each vertex in $V$.
We use these coordinates to represent vertices (i.e., instead of indices), hence,
\begin{align*}
    V \subset \mathcal{P} \left( \{(i,j) \in [0 \, \ldots \, R[\, \times \, [0 \, \ldots \, C[\} \right),
\end{align*}
where $R$ is the number of rows and $C$ is the number of columns of the grid and $\mathcal{P}(\cdot)$ denotes the power set.
To facilitate manipulation of coordinates, let $I_v$ (resp. $J_v$) be the set of rows (resp. columns) of a vertex $v$ in the two-dimensional grid:
\begin{align*}
& \forall v \in V, 
& I_v = \left\{ i \in [0 \, \ldots \, R[ \,\middle|\, \exists j \in [0 \, \ldots \, C[ \,\wedge\, (i,j) \in v \right\}, \\
& & J_v = \left\{ j \in [0 \, \ldots \, C[ \,\middle|\, \exists i \in [0 \, \ldots \, R[ \,\wedge\, (i,j) \in v \right\}.
\end{align*}
\paragraph{Tree indexation}
\label{paragraph:Tree_indexation}
To more compactly represent the data, we summarise fractions of the grid by a vertex $v$, i.e., a model $\mathcal{M}_v$, based on the covered data.
To cover the full information of the data, the tree graph must have as many leaves as the grid has elements.

We obtain a multi-scale representation, where all vertices of the tree are associated with a subset of the grid.
In our case, children are a nested split of the set of vertices covered by their parent.
The connection of parents to their children (i.e., parts and their nested partition) is represented by the edges of the tree.

\paragraph{Tree indexation algorithms}
\label{paragraph:Tree_indexation_algorithms}
For tree indexation, we use quad tree and binary tree algorithms.
In every step $t \in \mathbb{N}$ of the algorithm, the tree is updated.
At $t=0$, the tree $T_0$ consists of a single vertex $v_0$,
\begin{align*}
  T_0 = (V_0, E_0) \text{ with }
  \: V_0 = \{v_0\},
  \: E_0 = \emptyset,
  \: v_0 = \left \{ (i,j) \in [0 \, \ldots \, R[ \, \times \, [0 \, \ldots \, C[ \, \right \},
\end{align*}
which summarises the information of the whole grid (i.e., $\mathcal{M}_0$).

For \textbf{quad trees}, vertices are either leaves or have four children. 
We use a deterministic approach to divide a vertex into its children.
Thus, a part is subdivided into four mostly equal parts along the rows and columns:
\begin{align*}
	& \forall t \in \mathbb{N},
	& V_{t+1} & = V_t \, \underset{\forall v \in le(V_t)}{\mathrm{\cup}} \, ch(v),
	& ch(v) & = \left \{ v_{\squareulquad\quad\!\!\!\!\!\!\!}, v_{\squareurquad\quad\!\!\!\!\!\!\!}, v_{\squarellquad\quad\!\!\!\!\!\!\!}, v_{\squarelrquad\quad\!\!\!\!\!\!\!} \right \},
\end{align*}
\begin{align*}
	& \text{with } & v_{\squareulquad\quad\!\!\!\!\!\!\!} & = \left \{ (i,j) \in v \mid i < \left \lceil{ mid(I_v)} \right \rceil,  j < \left \lceil{mid(J_v)} \right \rceil \right \},\\
	& & v_{\squareurquad\quad\!\!\!\!\!\!\!} & = \left \{ (i,j) \in v \mid i < \left \lceil{mid(I_v)} \right \rceil,  j \geq \left \lceil{mid(J_v)} \right \rceil \right \},\\
	& & v_{\squarellquad\quad\!\!\!\!\!\!\!} & = \left \{ (i,j) \in v \mid i \geq \left \lceil{mid(I_v)} \right \rceil,  j < \left \lceil{mid(J_v)} \right \rceil \right \},\\
	& & v_{\squarelrquad\quad\!\!\!\!\!\!\!} & = \left \{ (i,j) \in v \mid i \geq \left \lceil{mid(I_v)} \right \rceil,  j \geq \left \lceil{mid(J_v)} \right \rceil \right \},
\end{align*}
where $mid(\bullet) = \frac{max(\bullet) - min(\bullet)}{2} + min(\bullet)$.\\
The tree is updated by the corresponding directed edges,
\begin{align*}
  & \forall t \in \mathbb{N},
  & E_{t+1} = E_t \, \underset{\forall v \in le(V_t)}{\mathrm{\cup}} \, \left \{ (v, v_{\squareulquad\quad\!\!\!\!\!\!\!}), (v, v_{\squareurquad\quad\!\!\!\!\!\!\!}), (v, v_{\squarellquad\quad\!\!\!\!\!\!\!}), (v, v_{\squarelrquad\quad\!\!\!\!\!\!\!}) \right \}.
\end{align*}
Tree indexation stops at $t_{max}$, where the tree consists of as many leaves as the grid has elements.
This implies that there is no leaf $v$ of $T_{t_{max}}$ that still has children.

For \textbf{binary trees}, vertices are either leaves or have two children.
We divide a vertex into its two children (i.e., a part into two, along the rows or along the columns) based on the data (i.e., a score).
We define
\begin{align*}
  \forall t \in \mathbb{N},
  \, \forall v \in & le(V_t),\\
  & \, \forall r \in I_v,
  \, v_{\squaretopblack\quad\!\!\!\!\!\!\!}^r = \left \{ (i,j) \in v \mid i < r \right \},
  v_{\squarebotblack\quad\!\!\!\!\!\!\!}^r = \left \{ (i,j) \in v \mid i \geq r \right \}
  = \overline{v_{\squaretopblack\quad\!\!\!\!\!\!\!}^r},\\
  & \, \forall c \in J_v,
  \, v_{\squareleftblack\quad\!\!\!\!\!\!\!}^c = \left \{ (i,j) \in v \mid j < c \right \},
   v_{\squarerightblack\quad\!\!\!\!\!\!\!}^c = \left \{ (i,j) \in v \mid j \geq c \right \}
  = \overline{v_{\squareleftblack\quad\!\!\!\!\!\!\!}^c}.
\end{align*}
For a possible division of a vertex $v$ into its children $\{ \langle v_{\squareleftblack\quad\!\!\!\!\!\!\!}^r, \overline{v_{\squareleftblack\quad\!\!\!\!\!\!\!}^r}\rangle \}_{r \in le(v)}$\\
(or $\{ \langle v_{\squaretopblack\quad\!\!\!\!\!\!\!}^c, \overline{v_{\squaretopblack\quad\!\!\!\!\!\!\!}^c}\rangle \}_{c \in le(v)}$) we can fit a separate model.
Thus, for each possible division, we can compute a score and among all possible partitions of each leaf at step $t \in \mathbb{N}$ select the best partition.
We will refer to the best possible partition as $\{ \langle v^\ast, \overline{v^\ast}\rangle \}_{\ast \in le(v)}$ in the following.

The vertices $\{ \langle v^\ast, \overline{v^\ast}\rangle \}_{\ast \in le(v)}$ and their corresponding edges $\{ \langle (v, v^\ast), (v, \overline{v^\ast})\rangle \}_{\ast \in le(v)}$ are used to update the set of vertices $V_t$ and edges $E_t$ of the tree $T_t$:
\begin{align*}
  & \forall t \in \mathbb{N},
  & V_t & \, \underset{\forall v \in le(V_t)}{\mathrm{\cup}} \, \left \{ \langle v^\ast, \overline{v^\ast} \rangle \right \},
  & E_t & \, \underset{\forall v \in le(V_t)}{\mathrm{\cup}} \, \left \{ \langle (v, v^\ast), (v, \overline{v^\ast}) \rangle \right \}.
\end{align*}
As for the quad tree, tree indexation stops at $t_{max}$.
$T_{t_{max}} = \left( V_{t_{max}}, E_{t_{max}} \right)$ will be referred to as $T = (V, E)$ in the following.

Note that we set the minimum number of rows or columns of a part covered by a vertex to be one.
In the case of \gls{grf}s, we can fit up to five parameters per part, consisting of expectation (i.e., $\mu$), variance (i.e., $\sigma$), and possibly correlation within rows and columns (i.e., $\rho_r$, $\rho_c$), and ratio of nugget parameter to residual variance (i.e., $\phi$).
To make this model estimable, we constrain the minimum number of plots covered by a vertex to be six.

\subsection{Identification}
\label{subsec:Identification}
We use an iterative approach to find the partition (i.e., subset of vertices of the tree) that represents the data best.
We denote this subset by $V_{id}$.
Its leaves $le(V_{id})$ correspond to the identified patches.
We denote the number of leaves by $m_{id}$ in the following.

We use two identification algorithms.
The \textbf{Top-Down algorithm} iteratively recovers parts from $v_0$ to the leaves (i.e., from the least to the most detailed representation of the data).
In every step it decides if a split of a vertex to its children is accepted.
The \textbf{Bottom-Up algorithm} iteratively recovers parts from the leaves to $v_0$ (i.e., from the most to the least detailed representation of the data).
In every step it decides if merges of vertices to their parents are accepted.
Note that the identified patches $le(V_{id})$ may be part of different levels of detail.

\subsection{Authentication}
\label{subsec:Authentication}
Using the identification algorithms described above,
we only verify if the model of a vertex is sufficiently different from the model of its sibling(s), ignoring all other collateral relatives (e.g., parent's siblings).
By construction, the algorithms generally tend to estimate too detailed partitions (i.e, too many patches $le(V_{id})$).
Therefore, we need to determine which identified patches can be better modelled by one instead of two models.

This is a again a combinatory problem, which we solve iteratively by successive merging.
At each step, we let us guide by a matrix of a posteriori probabilities $\Pi_{auth}$,
\begin{align*} 
& \forall (p, q) \in (1, \ldots, m)^2, 
& \Pi_{auth} = (\pi_{pq}) \propto P_{Z_{A_p} = q}(\boldsymbol{Y}_{A_p} = \boldsymbol{y}_{A_p}),
\end{align*} 
where the variable $Z_{A_p}$ indicates to which patch the patch $A_p$ is associated to, and $P_{Z_{A_p} = q}(\boldsymbol{Y}_{A_p} = \boldsymbol{y}_{A_p})$ describes the probability to observe the data covered by the patch $p$ under the constraint that is is merged to the data covered by the patch $q$.
In the first iteration, the patches correspond to the identified patches $le(V_{id})$ and  $Z_{A_p} = p \quad \forall p \in (1, \ldots, m_{id})$.

In each iteration, we merge those patches with the maximal $\pi_{pq}$ and estimate a score based on the model.
We stop when all patches are merged and choose the iteration with the best score to obtain the $m_{auth}$ merged patches.
Note that we can either constrain the authentication algorithm to merge only adjacent patches (i.e., local authentication), or use no constrain (i.e., global authentication).
For local authentication, we set the minimum number of plots per patch to two.
For global authentication, we set it to six in the case of \gls{grf}.

\subsection{Verification}
\label{subsec:Verification}
To obtain more realistic shapes of the detected patches, we first determine all bordering plots (i.e., all plots from a part that are direct neighbours of another part).
We then iteratively go through all bordering plots of all parts, and estimate the a posteriori probability for it to stay in its patch or to switch to its neighbouring patches,
\begin{align*}
	\forall (p, q) \in (1, \ldots, m_{auth})^2, \quad
	\forall (i, k) \in (1, \ldots, R)^2, \quad
	\forall (j, l) \in (1, \ldots, C)^2,\\
	\pi_{p_{ij}q_{kl}} \propto 
	\begin{cases}
		P_{Z_{y_{ij}}=p,\, Z_{y_{kl}}=q}(y_{ij}, y_{kl}) 
		& 
		\begin{aligned}
			&\text{if}
			& i - k \leq 1,\ j - l \leq 1, \text{ and } y_{ij} \neq y_{kl}
		\end{aligned} 
		\\[1ex]
		0 & \text{otherwise,}
	\end{cases}
\end{align*}
where the variable $Z_{y_{ij}}$ indicates to which patch the data of plot $y_{ij}$ is associated to, and $P_{Z_{y_{ij}}=p,\, Z_{y_{kl}}=q}(y_{ij}, y_{kl}) $ describes the probability to observe the data covered by the neighbouring plots $y_{ij}$ and $y_{kl}$ under the constraint that $y_{ij}$ is associated to patch $p$ and $y_{kl}$ is associated to patch $q$.
In the first iteration, the shape of the patches corresponds to the patches found by the authentication.\\
This gives us the a posteriori probability of all possible switches.
We switch the plot with highest a posteriori probability $\pi_{p_{ij}q_{kl}}$ to the corresponding neighbouring patch.
The algorithm stops when switching of plots does not show higher a posteriori probabilities than staying in the current part.
Note that we re-estimate the  model of the newly shaped patches at each iteration.
To avoid convergence problems, we do not use model selection.
For instance, if the previous patch was estimated to be of type \gls{ar1}, we re-estimate all parameters of \gls{ar1} with the added plot.
In a last step, we estimate the model of each residual patch using model selection.

The verification algorithm may lead to a separation of a patch into disjoint sub-areas.
For the first cycle of authentication and verification (i.e., local authentication), we set the minimum number of plots per sub-area to two (i.e., similarly as for local authentication).
To ensure that the resulting separated areas are not too small, we set the minimum number of plots per sub-area to six in the case of \gls{grf} for the second cycle of authentication and verification (i.e., global authentication).

\section{Simulation Study}
\label{sec:Simulation}
\subsection{Simulation of residual patches in two-dimensional grids}
\label{subsec:Simulation}
\begin{figure}[H]
	\centering
	\begin{subfigure}{0.2\textwidth} 
    		\centering
    		\includegraphics[width=\textwidth]{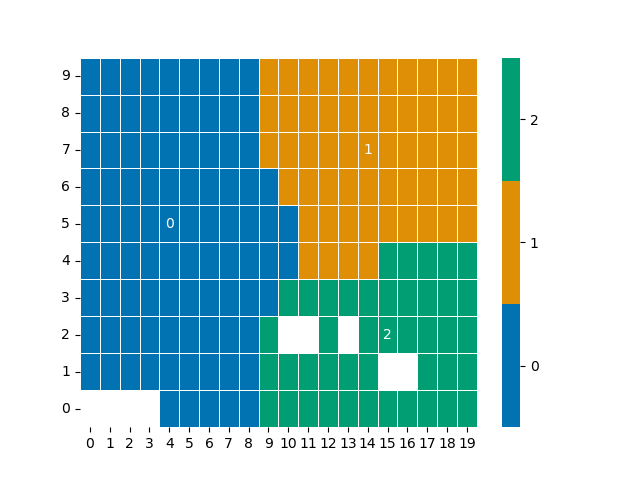}
 		\caption[Sufigure 2a]{\label{subfig:2a}}
	\end{subfigure} 
	\begin{subfigure}{0.2\textwidth} 
    		\centering
    		\includegraphics[width=\textwidth]{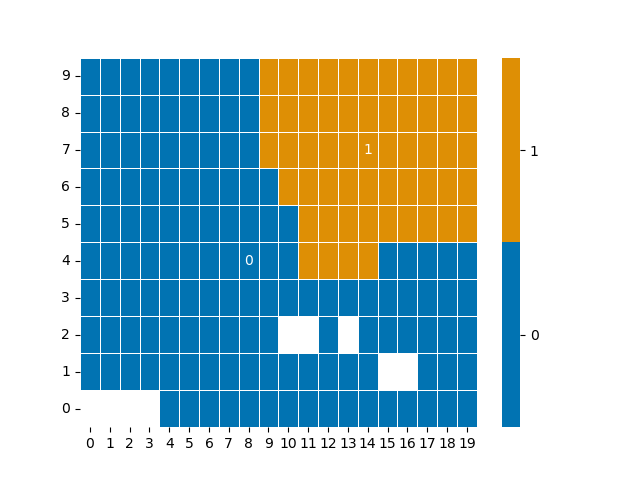}
 		\caption[Sufigure 2a]{\label{subfig:2b}}
	\end{subfigure} 
	\begin{subfigure}{0.2\textwidth} 
    		\centering
    		\includegraphics[width=\textwidth]{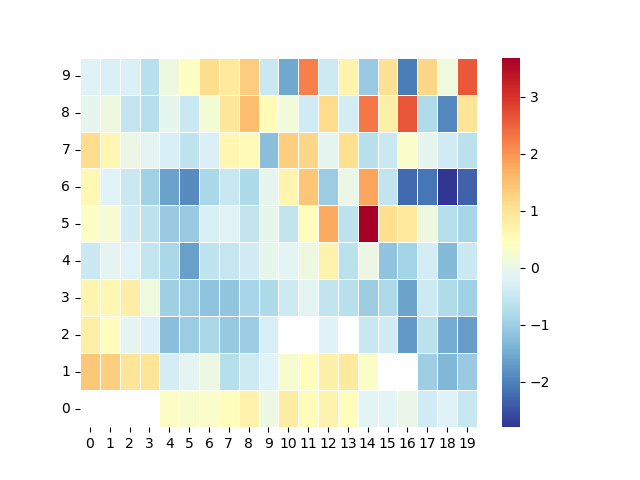}
 		\caption[Sufigure 2c]{\label{subfig:2c}}
	\end{subfigure} 
	\begin{subfigure}{0.2\textwidth} 
    		\centering
    		\includegraphics[width=\textwidth]{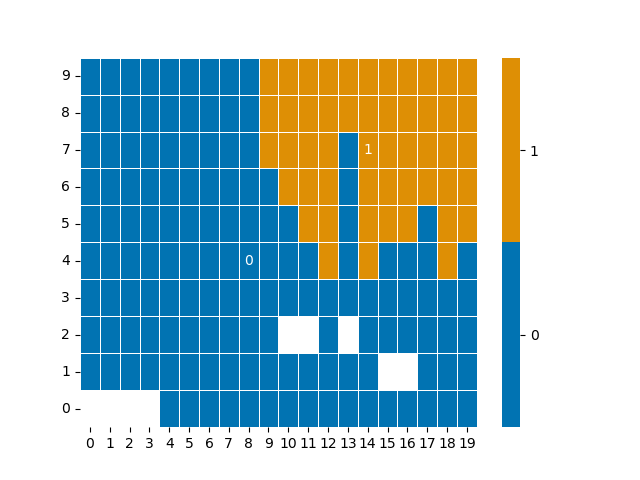}
 		\caption[Sufigure 2d]{\label{subfig:2d}}
	\end{subfigure} 
    \caption[Figure 2]{\label{fig:2}
    Simulation of residual patches: Simulation of three patches using Lloyd algorithm (a);
    Merging of two patches (b);
    Residuals drawn from a distinct Gaussian Random Field within each of the patches (c);
    Verification of the shape of each residual patch (d).
    }
\end{figure}

We conduct a simulation study to test the performance of the proposed quality control method.
More precisely, we want to test how well the method can reliably detect nonstationarity in agricultural field trials (i.e., number, location, and shape of detected residual patches).
We therefore simulate residual patches in two-dimensional grids that lead to the problem of nonstationarity in both mean and variance-covariance.
To ensure applicability of the method in real field trials, all simulations are based on historical field trial data.
An overview of the simulation of residual patches is shown in Figure \ref{fig:2}.

\begin{figure}[H]
	\centering
	\hspace{0.06\textwidth}
	\begin{subfigure}{0.25\textwidth} 
    		\centering
    		\includegraphics[width=\textwidth]{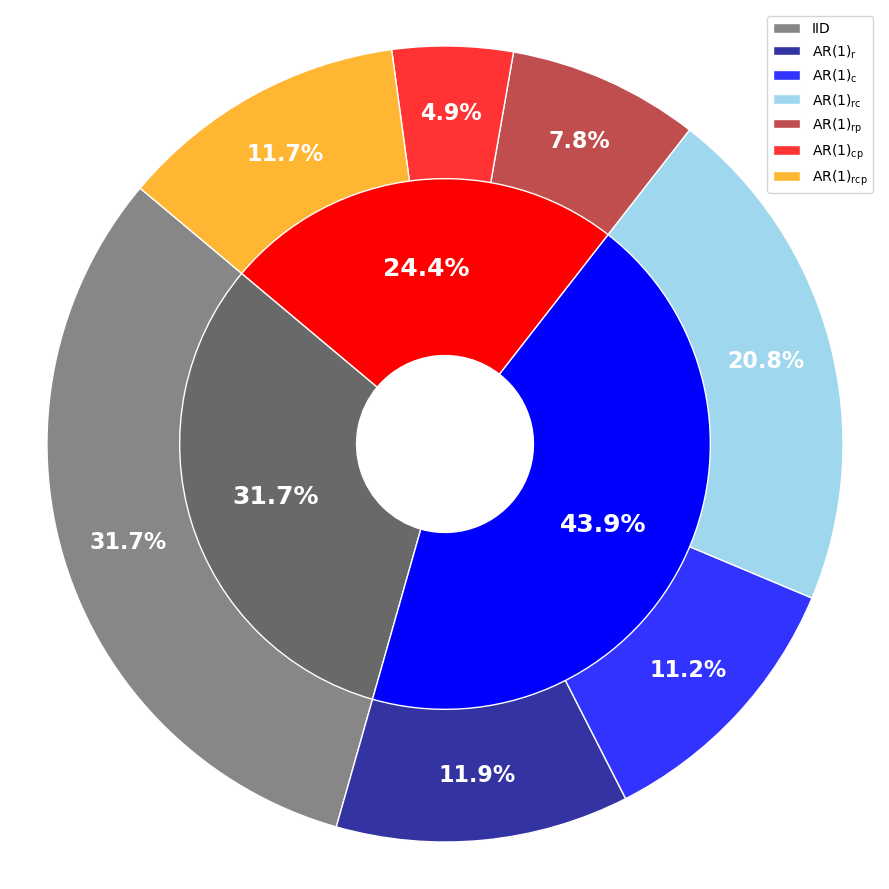}
 		\caption[Figure 3a]{\label{subfig:3a}}
	\end{subfigure} 
	\hspace{0.06\textwidth}
	\begin{subfigure}{0.33\textwidth} 
    		\centering
    		\includegraphics[width=\textwidth]{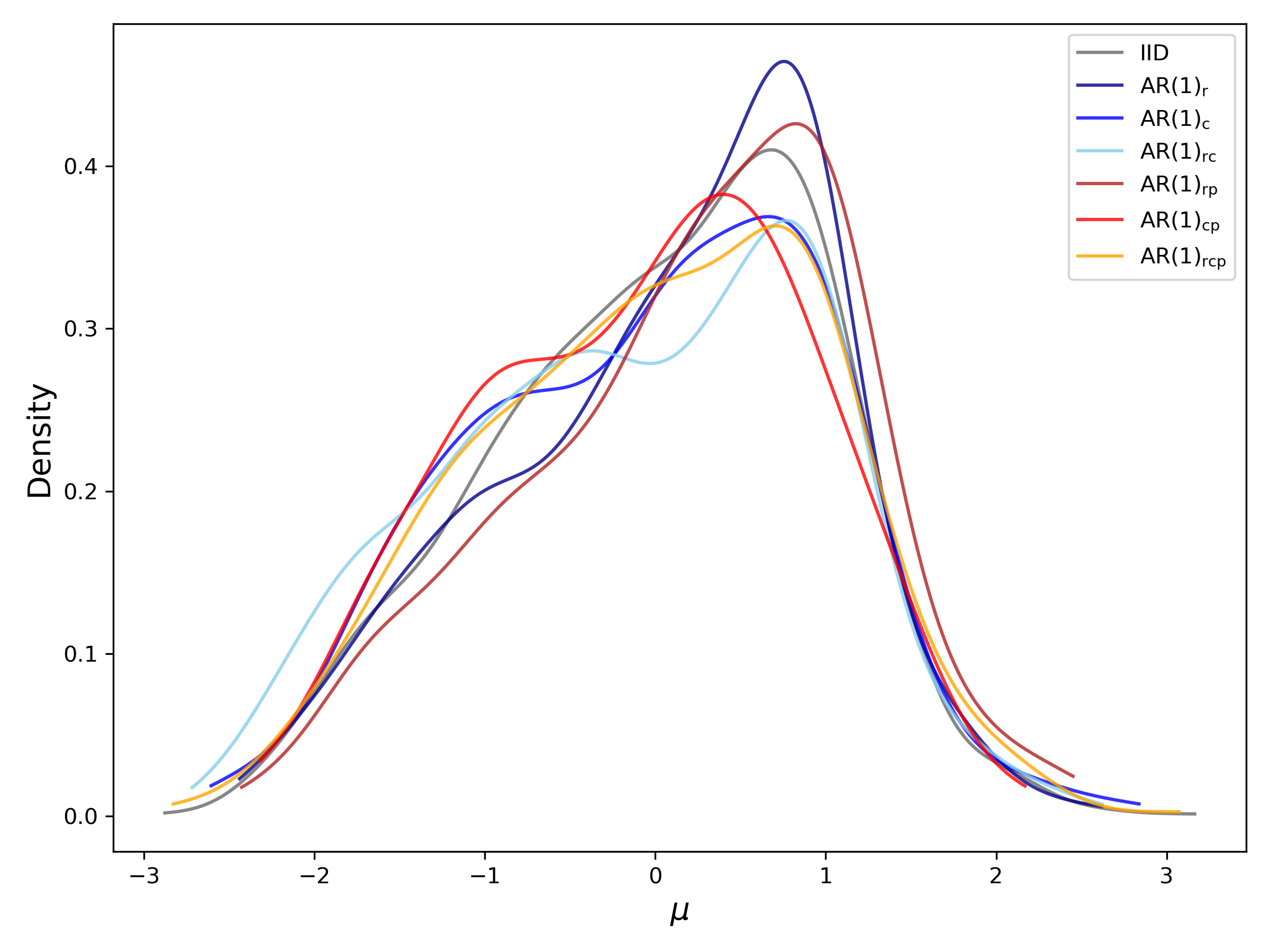}
 		\caption[Figure 3b]{\label{subfig:3b}}
	\end{subfigure} 
	\begin{subfigure}{0.33\textwidth} 
    		\centering
    		\includegraphics[width=\textwidth]{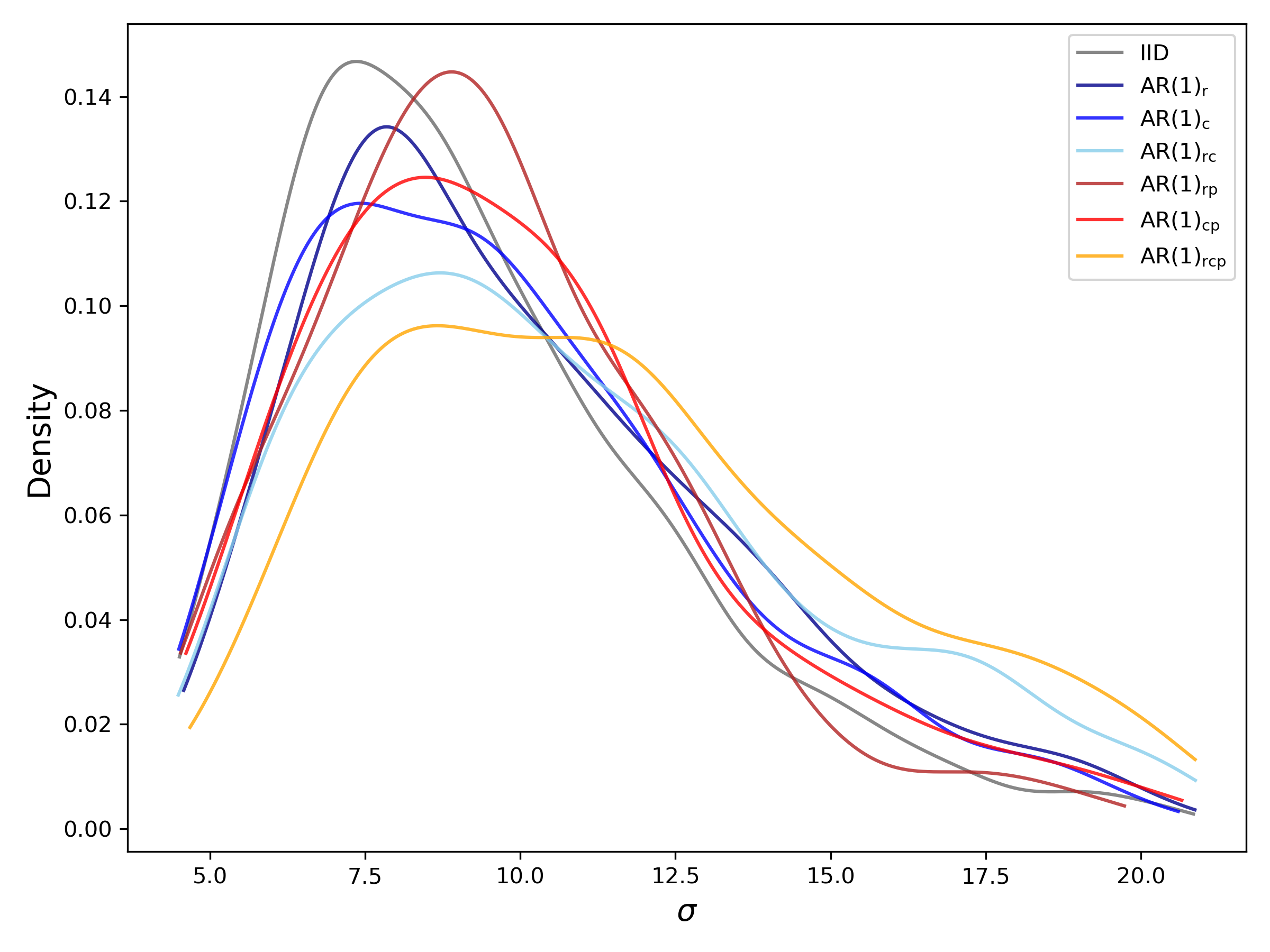}
 		\caption[Figure 3c]{\label{subfig:3c}}
	\end{subfigure} 
	\begin{subfigure}{0.33\textwidth} 
    		\centering
    		\includegraphics[width=\textwidth]{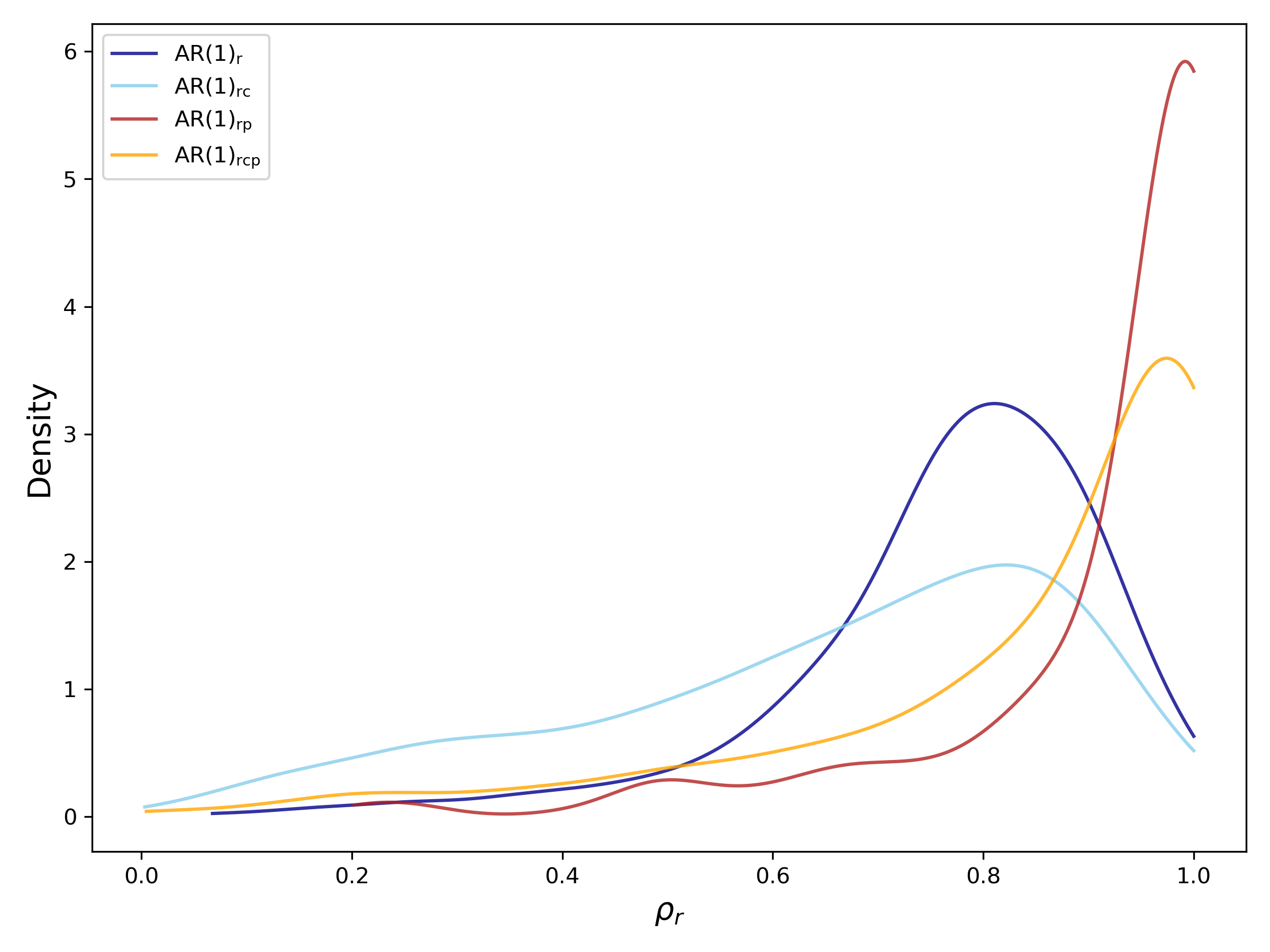}
 		\caption[Figure 3d]{\label{subfig:3d}}
	\end{subfigure} 
	\begin{subfigure}{0.33\textwidth} 
    		\centering
    		\includegraphics[width=\textwidth]{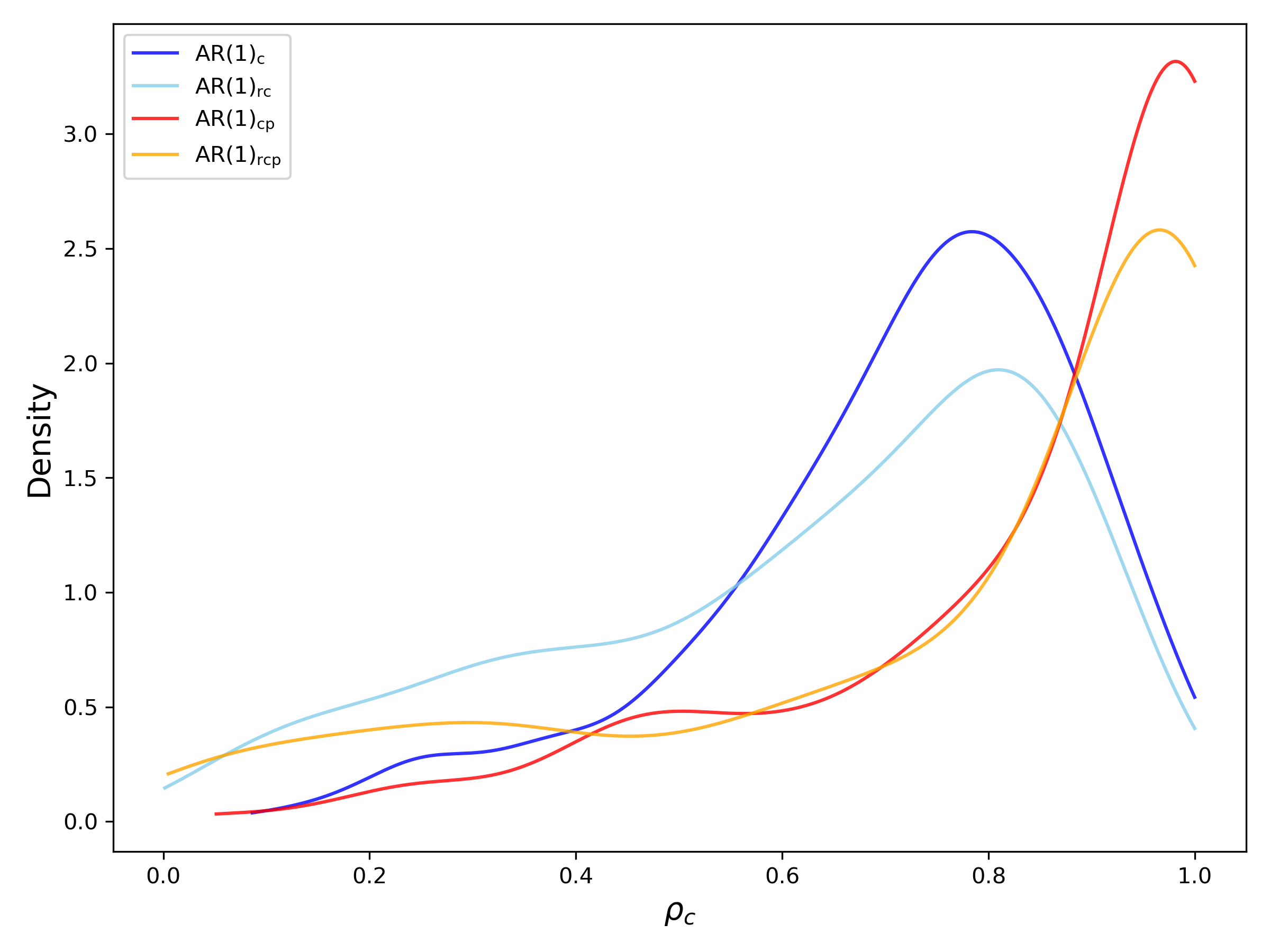}
 		\caption[Figure 3e]{\label{subfig:3e}}
	\end{subfigure} 
	\begin{subfigure}{0.33\textwidth} 
    		\centering
    		\includegraphics[width=\textwidth]{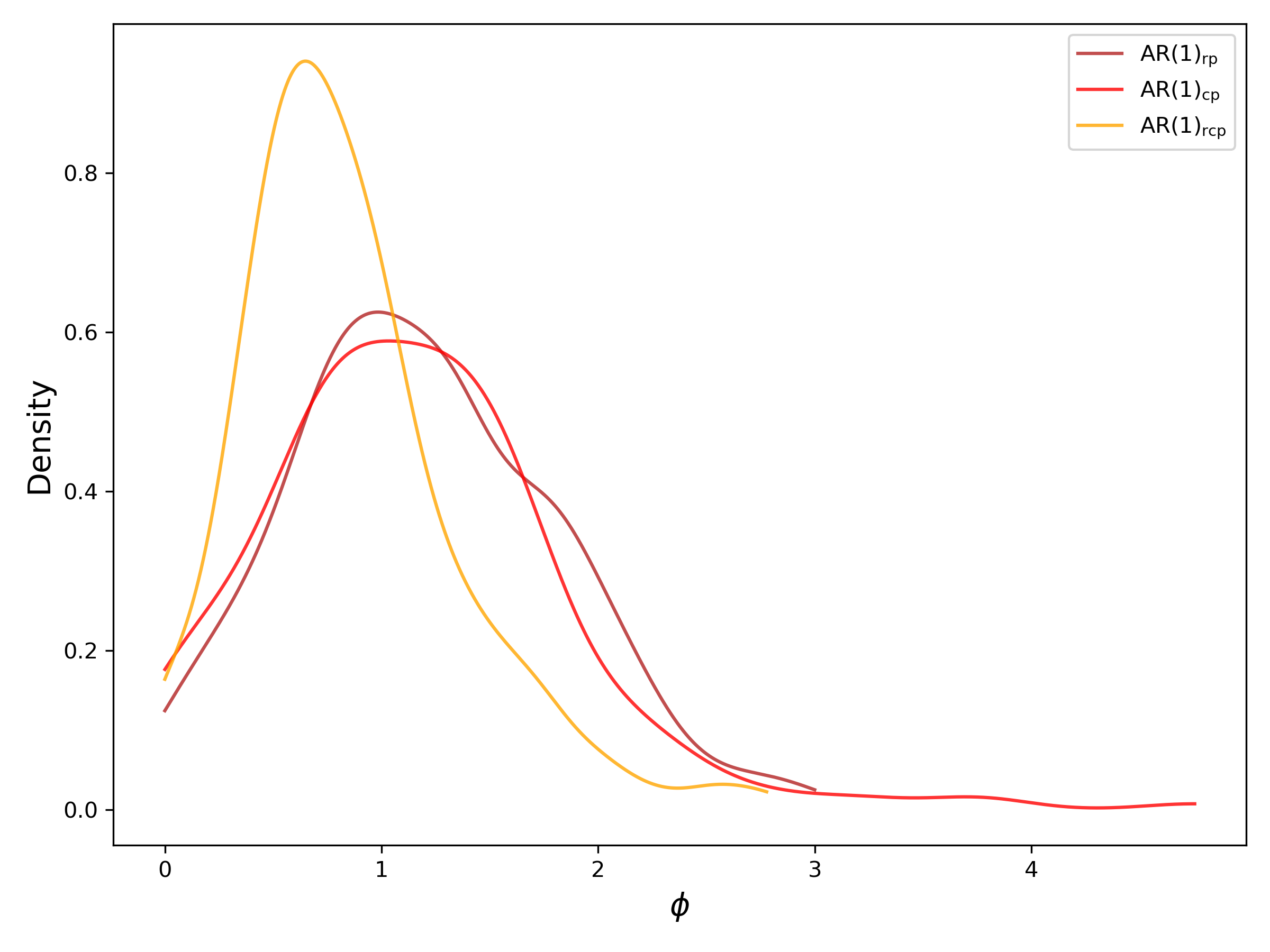}
 		\caption[Figure 3f]{\label{subfig:3f}}
	\end{subfigure} 
	\caption[Simulation of 1 residual patch]{\label{fig:3} (a) Abundance of Gaussian Random Field (GRF) types in over 4,000 historical field trials.
	GRF types are IID and AR(1) with correlations within rows and columns (indicated by the subscripts r and c) and nugget effect (indicated by the subscript p).
	Distribution of estimated
	trial mean $\mu$ (b),
	residual variance $\sigma^2$ (i.e., $\sigma$) (c),
	correlations within rows $\rho_r$ (d),
	correlations within columns $\rho_c$ (e),
	and ratio of nugget parameter to residual variance $\phi$ (f).
	Trial means are rescaled (i.e., z-standardisation).
	}
\end{figure}

\begin{enumerate}
\item For the \textbf{structure of two-dimensional grids} (i.e., rows, columns, missing data), we use existing field plan structures from historical field trials.
\item For the simulation of \textbf{position, size, and shape of the patches} within the field, we use the Llyod algorithm with euclidean distance (see \cite{lloyd1982least}).
To obtain realistic shapes of the patches (i.e., convex and concave shapes), we first simulate three patches (see Figure \ref{subfig:2a}) and then merge two of them (see Figure \ref{subfig:2b}).
\item Within each of the two patches, \textbf{residuals} are drawn from distinct \gls{grf}s (see Figure \ref{subfig:2c}).
We draw the parameters for the definition of each \gls{grf} from estimated parameters of several thousands historical field trials.
For each historical trial, we obtain the estimates for trial mean $\mu$, residual variance $\sigma^2$ (i.e., $\sigma$), correlations within rows and columns $\rho_r$ and $\rho_c$, and the ratio of nugget parameter to residual variance $\phi$.
For more information about the distribution of each estimated parameter see Figure \ref{fig:3}.
\item We use the authentication algorithm described in Section \ref{subsec:Authentication} to ensure that we simulate two sufficiently distinct \textbf{residual patches} (i.e., that result in the problem of nonstationarity).
If the residual patches are too similar and merged to a single patch, we simulate again.
\item Although the residuals within each patch are drawn from distinct \gls{grf}s, the simulated data might suggest a different shape of the residual patches than the initially simulated patches.
Therefore, we use the verification algorithm described in Section \ref{subsec:Verification} to rechallenge the borders of the simulated residual patches.
This gives us the \textbf{expected shape of residual patches} (i.e., oracle) (see Figure \ref{subfig:2d}).
\end{enumerate}

We set the minimum number of plots per simulated residual patch to eight.
In total, we simulate 100 trials.

\subsection{Evaluation of tree indexation}
\label{subsec:Evaluation_tree_indexation}
We want to determine whether using tree graphs to detect residual patches is a valid approach and, if so, to identify the most suitable tree.
To do this, we apply the tree algorithms and replace all subsequent steps by oracle algorithms, where possible.
This ensures that we can specifically focus on the potential of tree graphs. 

For tree indexation, we use the data-driven binary tree and the deterministic quad tree as described in Section \ref{subsec:Tree_indexation}.
For identification, we use the Bottom-Up algorithm as described in Section \ref{subsec:Identification}.
However, to decide if a merge should be accepted, we do not use scores computed based on the data.
Instead, we compute the similarity between the simulated residual patches and a vertex or its children.
We use the Bottom-Up algorithm because, in contrast to the Top-Down algorithm, it starts at greater depths and thus has the tendency to detect more patches.
If optimally merged, using the Bottom-Up algorithm leads to a higher chance of detecting the simulated residual patches.
It is therefore the best choice for the evaluation of tree indexation.
To decide which patches should be merged, we use the global authentication algorithm as described in Section \ref{subsec:Authentication}.
However, instead of  calculating $\Pi_{auth}$ (i.e., matrix of a posteriori probabilities), we use the similarity between the simulated residual patches and all possible merges.
In each iteration, we determine which merge results in highest similarity to the simulated residual patches.
We iterate until all patches are merged and then choose the iteration with highest similarity to the simulated residual patches.
We use the verification algorithm as described in Section \ref{subsec:Verification}.
Note that for tree evaluation, we only use one single round of global authentication followed by one single round of verification.

\begin{figure}[H]
    \centering
	\begin{subfigure}{0.4\textwidth} 
    		\centering
    		\includegraphics[width=\textwidth]{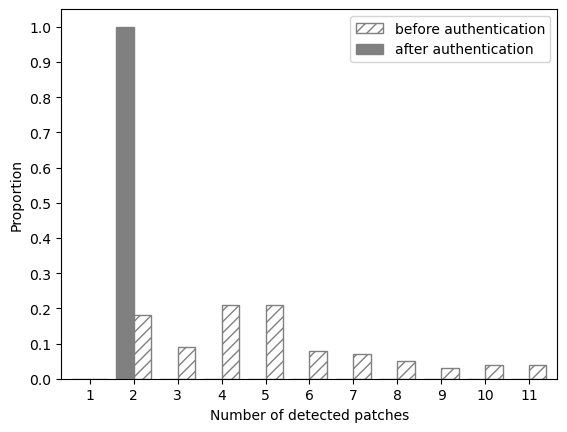}
    		\caption[Subfigure 4a]{\label{subfig:4a}}
	\end{subfigure}
	\begin{subfigure}{0.4\textwidth} 
    		\centering
    		\includegraphics[width=\textwidth]{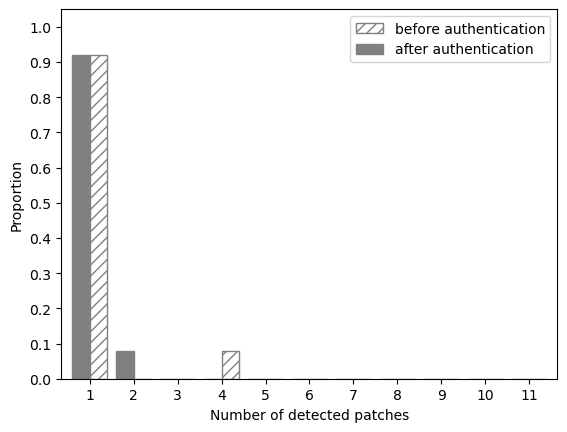}
    		\caption[Subfigure 4a]{\label{subfig:4b}}
	\end{subfigure}
    \caption[Figure 4]{\label{fig:4}
    Reduction of detected patches through authentication for (a) binary tree and (b) quad tree.
    Authentication and precedent identification based on oracle algorithms.
    }
\end{figure}

\begin{figure}[H]
	\centering
	\includegraphics[width=0.45\textwidth]{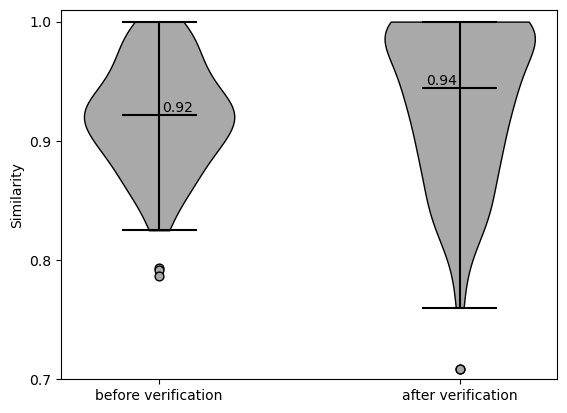}
 	\caption[Figure 5]{\label{fig:5}
 	Improvement of the shape of detected residual patches through verification.
 	Precedent tree indexation using data-driven binary tree; identification and authentication based on oracle algorithms.
    }
\end{figure}

We determine data-driven binary tree as the most suitable tree graph.
Its partitions provide a good basis for the detection of stationary areas as oracle identification and subsequent oracle authentication lead to the estimation of mostly two patches (see Figure \ref{subfig:4a}).
Their similarity to the simulated residual patches is high and can be slightly improved by subsequent verification (see Figure \ref{fig:5}).
The deterministic quad tree, on the other hand, mostly leads to the identification of only one patch (see Figure \ref{subfig:4b}).
It is therefore not flexible enough for the detection of residual patches.
In the following, we therefore continue with binary tree.

\subsection{Evaluation of identification}
\label{subsec:Evaluation_identification}
In the previous Section \ref{subsec:Evaluation_tree_indexation}, we determined the data-driven binary tree as the best tree.
The obtained similarity using the oracle algorithms is a reference for the maximal performance that we can expect of the entire method.
Now, we want to evaluate the actual performance of the proposed method as described in Section \ref{sec:Proposed_Method}.
More specifically, we focus on the difference in performance of Top-Down and Bottom-Up identification.
We thus apply the proposed method using the best tree (i.e., binary tree) with subsequent identification (i.e., Top-Down and Bottom-Up), authentication, and verification.

\begin{figure}[H]
    \centering
	\begin{subfigure}{0.32\textwidth} 
    		\centering
    		\includegraphics[width=\textwidth]{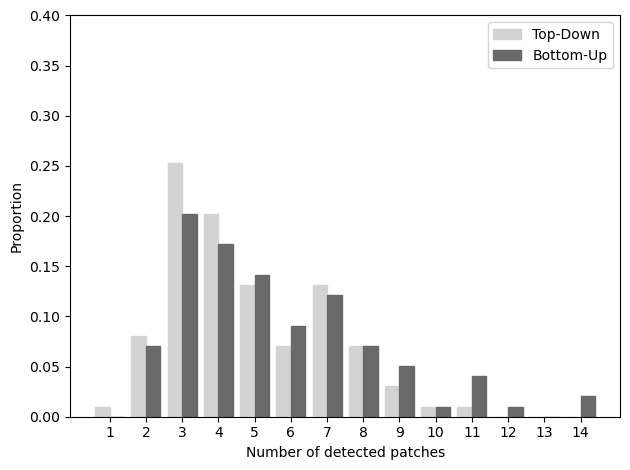}
    		\caption[Subfigure 6a]{\label{subfig:6a}}
	\end{subfigure}
	\begin{subfigure}{0.32\textwidth} 
    		\centering
    		\includegraphics[width=\textwidth]{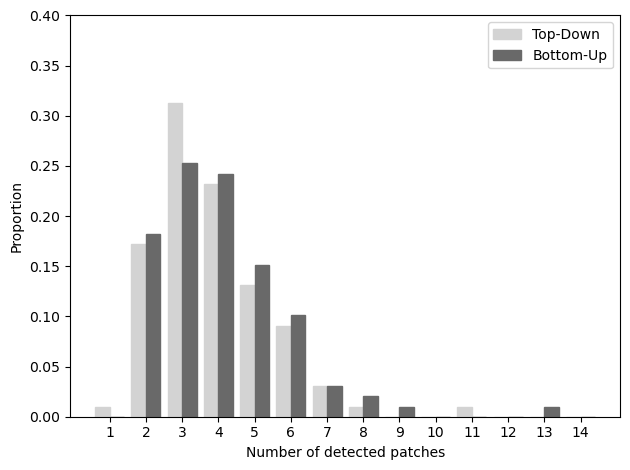}
    		\caption[Subfigure 6b]{\label{subfig:6b}}
	\end{subfigure}
	\begin{subfigure}{0.32\textwidth} 
    		\centering
    		\includegraphics[width=\textwidth]{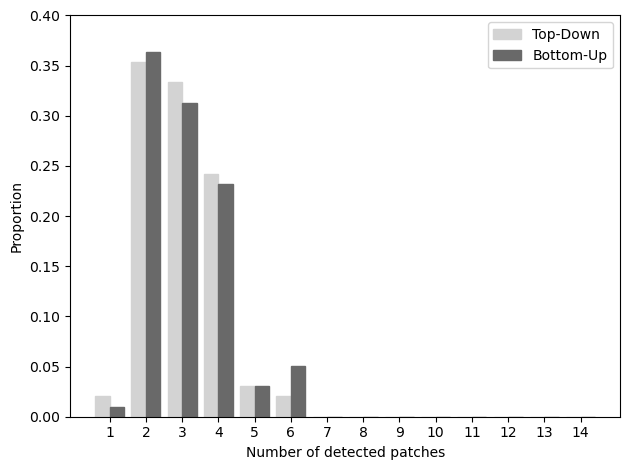}
    		\caption[Subfigure 6c]{\label{subfig:6c}}
	\end{subfigure}
    \caption[Figure 6]{\label{Eval_id_nopatches}
 	Number of detected patches for Bottom-Up and Top-Down algorithm after (a) identification, (b) local authentication, and (c) global authentication.
   }
\end{figure}

\begin{figure}[H]
    \centering
	\begin{subfigure}{0.45\textwidth} 
    		\centering
    		\includegraphics[width=\textwidth]{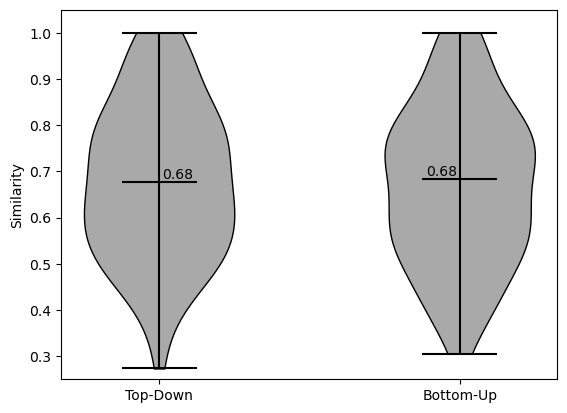}
    		\caption[Subfigure 7a]{\label{subfig:7a}}
	\end{subfigure}
	\begin{subfigure}{0.45\textwidth} 
    		\centering
    		\includegraphics[width=\textwidth]{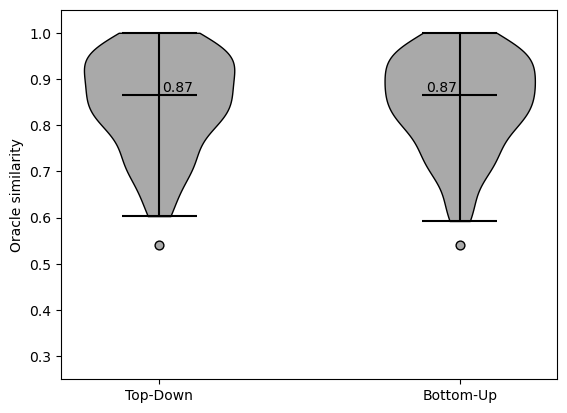}
    		\caption[Subfigure 7b]{\label{subfig:7b}}
	\end{subfigure}
    \caption[Figure 7]{\label{fig:7}
    Comparison of performance between Top-Down and Bottom-Up algorithm.
    Similarity (a) and improved similarity (b) through optimal merging.
   }
\end{figure}

As expected, Bottom-Up algorithm leads to the identification of more patches than Top-Down algorithm (see Figure \ref{subfig:6a}).
For both algorithms, the identified patches are successively merged by local and global authentication.
This results in the detection of one to six patches after global authentication (see Figure \ref{subfig:6b} and \ref{subfig:6c}).
Both identification algorithms show similar performance (see Figure \ref{subfig:7a}).
When more than two patches are detected, they are mostly subsets of the simulated patches and could be further merged.
Under the assumption of optimal merging, the results can be largely improved (see Figure \ref{subfig:7b}).
In the following, we continue with Top-Down algorithm since it is advantageous for optimising the workflow of the proposed method (see Section \ref{sec:Discussion}).

\subsection{Evaluation of authentication}
\label{subsec:Evaluation_authentication}
From the previous results in Section \ref{subsec:Evaluation_identification}, we conclude that authentication is not stringent enough and results could be improved through stricter merging.
We therefore apply a last cycle of global authentication to merge the detected patches further.
We control the level of stringency of authentication by setting different thresholds (e.g., 'anecdotal', 'decisive').

\begin{figure}[H]
\centering
\begin{minipage}{0.4\textwidth}
    \centering
    \includegraphics[width=0.9\textwidth]{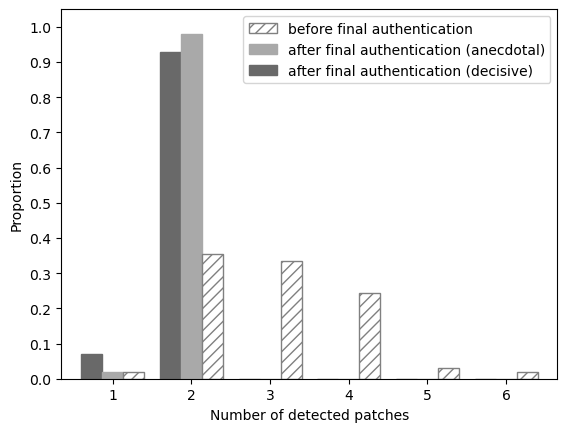}
    \caption[Figure 8]
    {\label{fig:8}
    Decreasing number of detected patches with increasing strength of merging.}
\end{minipage}
\hspace{1cm}
\begin{minipage}{0.4\textwidth}
    \centering
    \includegraphics[width=0.9\textwidth]{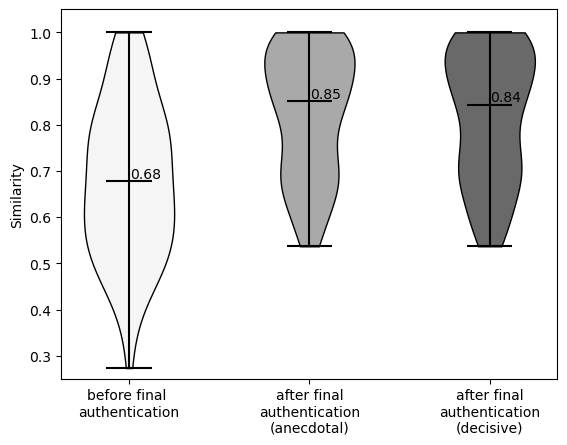}
    \caption[Figure 9]
    {\label{fig:9}
    Improved similarity of detected to simulated residual patches for different thresholds.}
\end{minipage}
\end{figure}

Increasing thresholds leads to a decreasing number of detected patches.
For our data, final authentication results in an estimation of mostly two residual patches with only small differences for different thresholds (see Figure \ref{fig:8}).
The similarity to the simulated residual patches is much higher than before final authentication and very similar among different thresholds (see Figure \ref{fig:9}).

\section{Real Data Illustration}
\label{sec:Real_Data_Illustration}
We test the proposed method on a disease trial data set from plant breeding practice.
Particularly in disease trials, we may assume unexpected field effects:
In disease trials, disease pressure plays an important, additional role as field effect.
Without disease pressure, the trait of interest (i.e., disease resistance) cannot be assessed.
Disease pressure thus has a strong, direct influence on the response (e.g., scoring values).
However, it may be very hard to control, particularly in trials with natural infection (e.g., through vectors such as insects).
But even with artificial contamination, disease pressure may depend on many other factors, that are not constant for all plots (e.g., date of contamination depends on the developmental stage of the plant; different temperatures for different contamination dates) \citep{{russell1978plant}, hickey2012rapid}.
We therefore consider the provided disease trial data set a very useful example, as it clearly illustrates the importance of accurate estimation in field trials and the need for stationarity verification.

\paragraph{Data}
Wheat varieties were tested for \gls{bydv} resistance.
\gls{bydv} is a disease that can affect barley, but also wheat and other cereals.
The virus is transmitted by aphids.
Symptoms of affected plants are small growth (i.e., dwarfism) and yellow leaves \citep{miller1997barley}.

\begin{figure}[H]
\centering
\begin{minipage}{0.36\textwidth}
	\vspace*{\fill}
    \centering
	\begin{subfigure}{0.48\textwidth} 
    		\centering
    		\includegraphics[width=\textwidth]{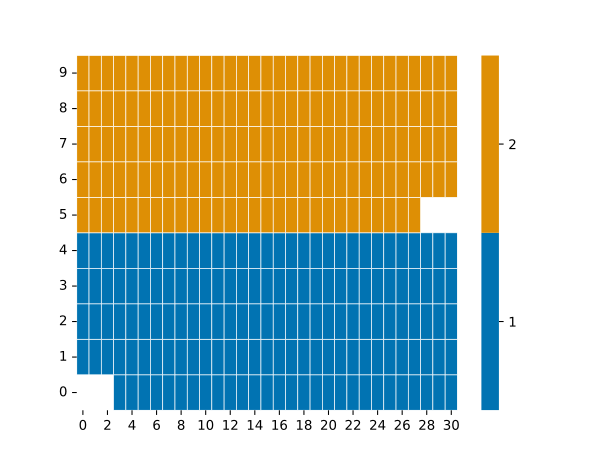}
    		\caption[Subfigure 10 a]{\label{subfig:10a}}
	\end{subfigure}
	\begin{subfigure}{0.48\textwidth} 
    		\centering
    		\includegraphics[width=\textwidth]{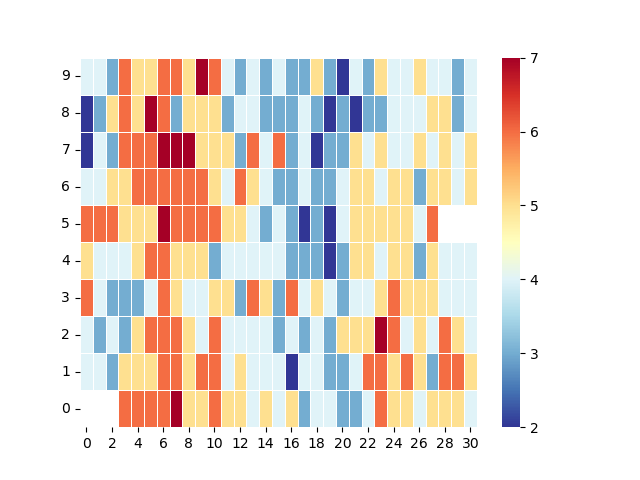}
    		\caption[Subfigure 10 b]{\label{subfig:10b}}
	\end{subfigure}
	\vspace*{\fill}
\end{minipage}
\begin{minipage}{0.5\textwidth}
    \begin{subfigure}{0.32\textwidth}
        \centering
        \includegraphics[width=\textwidth]{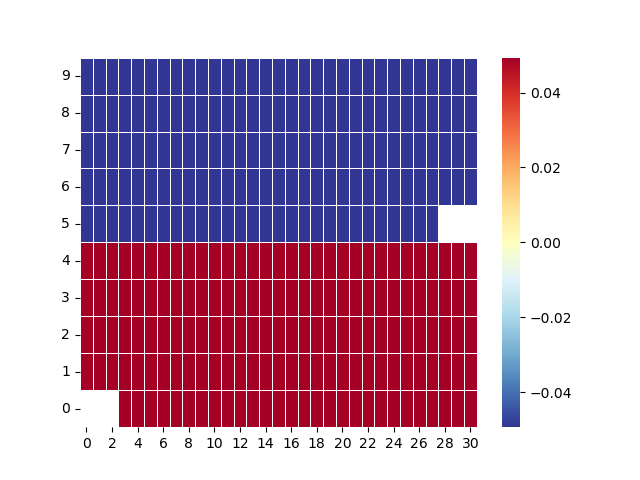}
        \caption[Subfigure 10 c]{\label{subfig:10c}}
    \end{subfigure}
    \begin{subfigure}{0.32\textwidth}
        \centering
        \includegraphics[width=\textwidth]{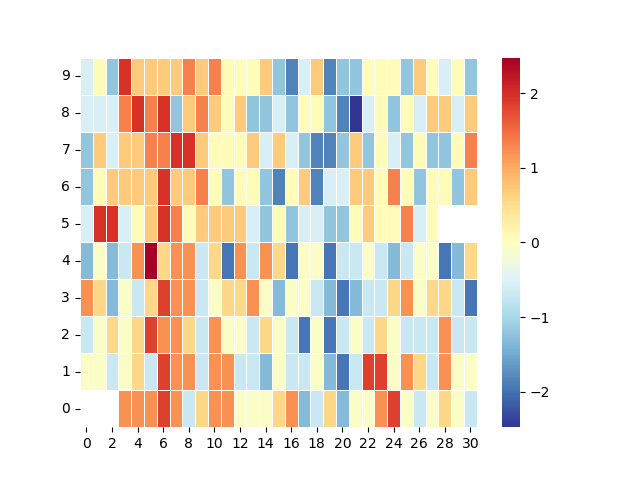}
        \caption[Subfigure 10 d]{\label{subfig:10d}}
    \end{subfigure}
    \begin{subfigure}{0.32\textwidth}
        \centering
        \includegraphics[width=\textwidth]{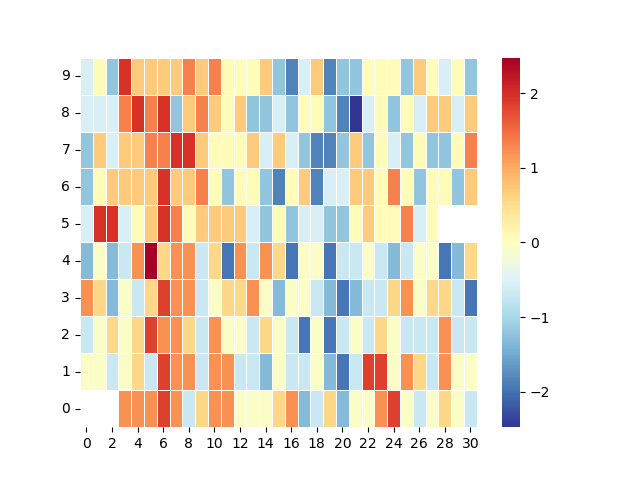}
        \caption[Subfigure 10 e]{\label{subfig:10e}}
    \end{subfigure}\\
    \begin{subfigure}{0.32\textwidth}
        \centering
        \includegraphics[width=\textwidth]{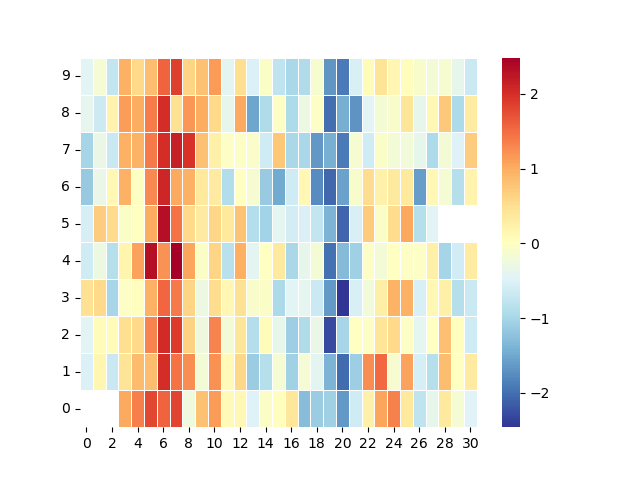}
        \caption[Subfigure 10 f]{\label{subfig:10f}}
    \end{subfigure}
    \begin{subfigure}{0.32\textwidth}
        \centering
        \includegraphics[width=\textwidth]{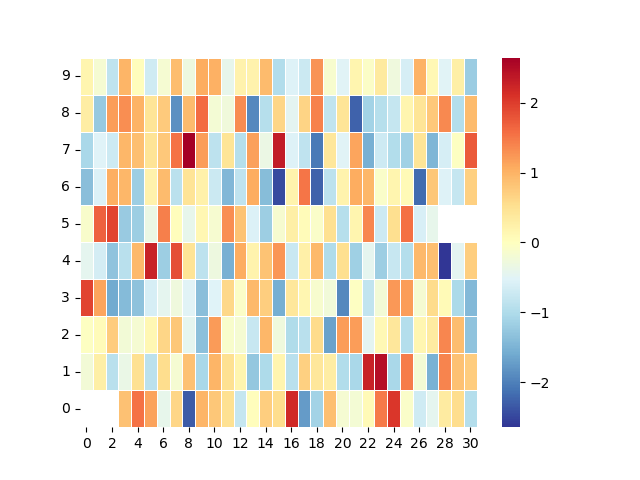}
        \caption[Subfigure 10 g]{\label{subfig:10g}}
    \end{subfigure}
    \begin{subfigure}{0.32\textwidth}
        \centering
        \includegraphics[width=\textwidth]{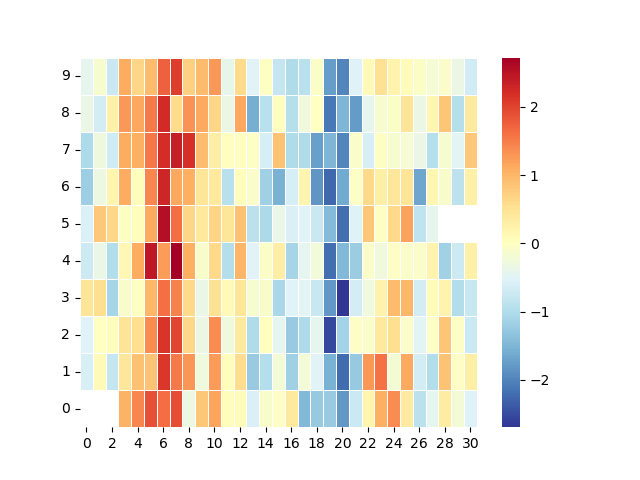}
        \caption[Subfigure 10 h]{\label{subfig:10h}}
    \end{subfigure}
\end{minipage}
    \caption[Figure 10]{\label{fig:10}
    Analysis of a real-data example.
    Blocking structure of the experimental design (a) and phenotypic data (b).
    Estimated field effect (c, f), 
    conditional residuals for manual quality control (d, g),
    and marginal residuals for automatic quality control (e, h).
    Analysis based on a linear model including block effects and IID (c, d, e) or AR(1) (f, g, h) variance-covariance structure.
    }
\end{figure}

The field trial was set up as a \gls{rcbd} in ten rows and thirty-one columns.
152 varieties were completely randomised within each of the two complete blocks.
Note that three plots at the corners of each block appear to be missing; this is simply because the number of varieties does not allow for perfectly rectangular blocks (see Figure \ref{subfig:10a}).
For each plot, a scoring value was estimated by a phenotyper using a 1-8 scale, where ordinal scores correspond to the discretisation of a metric, underlying disease severity scale (e.g., small values for low disease severity, high values for high disease severity) (see Figure \ref{subfig:10b}).

\paragraph{Analysis}
The data has been initially analysed by the breeders using a linear model including block effects and \gls{iid} variance-covariance structure.
No other effects were assumed and thus no model selection was performed, for example to confirm the type of \gls{grf} (i.e., assumption of mean and variance nonstationarity).

In plant breeding practice, disease scores are often treated as metric, assuming approximate normality of the response \citep{thoni1985auswertung, laidig2021breeding}.
The resulting estimated field effect, conditional, and marginal residuals are visualised in Figure \ref{subfig:10c}, \ref{subfig:10d}, and \ref{subfig:10e}.

\paragraph{Quality Control}
For manual quality control, the plant breeders used the estimated conditional residuals.
Visualisation as a heatmap indicates spatial patterns.
For instance, the residuals in the left part of the field appear to be higher than those in the right part.
Furthermore, there seems to be a strong level of autocorrelation (see Figure \ref{subfig:10d}).

\begin{figure}[H]
    \centering
	\begin{subfigure}{0.17\textwidth} 
    		\centering
    		\includegraphics[width=\textwidth]{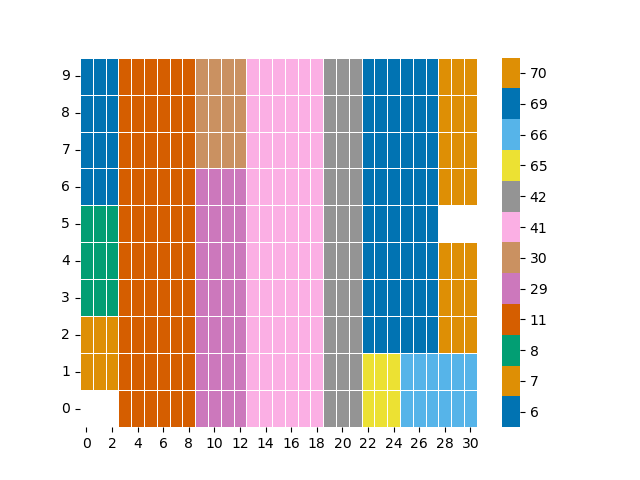}
    		\caption[Subfigure 11a]{\label{subfig:11a}}
	\end{subfigure} 
	\begin{subfigure}{0.17\textwidth} 
    		\centering
    		\includegraphics[width=\textwidth]{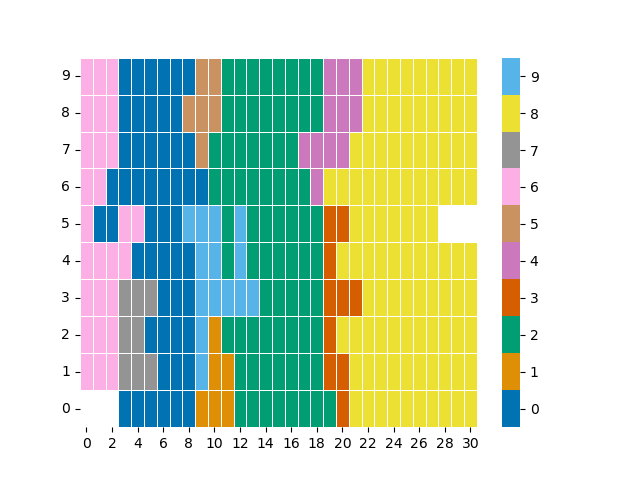}
    		\caption[Subfigure 11b]{\label{subfig:11b}}
	\end{subfigure}
	\begin{subfigure}{0.17\textwidth} 
    		\centering
    		\includegraphics[width=\textwidth]{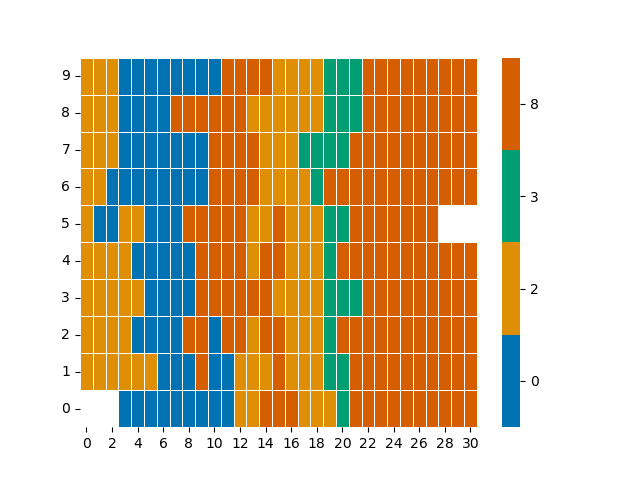}
    		\caption[Subfigure 11c]{\label{subfig:11c}}
	\end{subfigure}
	\begin{subfigure}{0.17\textwidth} 
    		\centering
    		\includegraphics[width=\textwidth]{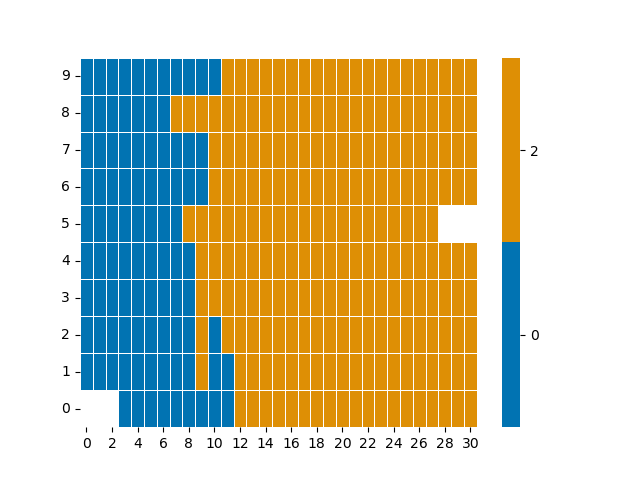}
    		\caption[Subfigure 11d]{\label{subfig:11d}}
	\end{subfigure}
	\begin{subfigure}{0.17\textwidth}
    		\centering
    		\rule{0pt}{0.8\textwidth}
    	\end{subfigure}\\
	\begin{subfigure}{0.17\textwidth} 
    		\centering
    		\includegraphics[width=\textwidth]{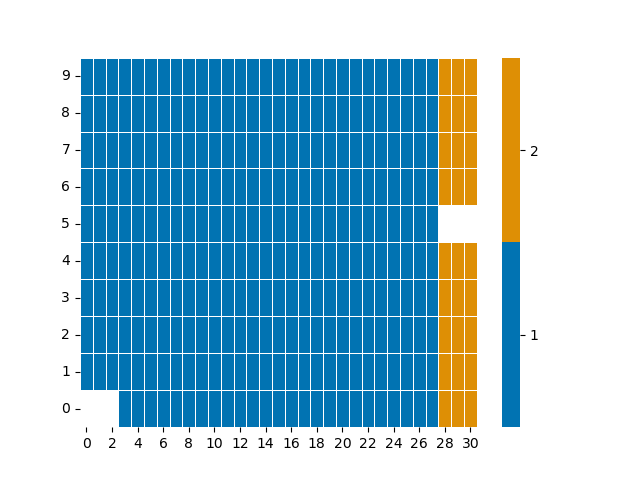}
    		\caption[Subfigure 11e]{\label{subfig:11e}}
	\end{subfigure}
	\begin{subfigure}{0.17\textwidth} 
    		\centering
    		\includegraphics[width=\textwidth]{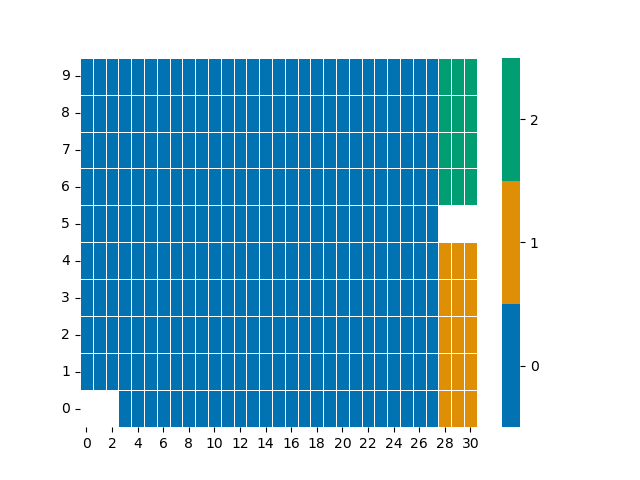}
    		\caption[Subfigure 11f]{\label{subfig:11f}}
	\end{subfigure}
	\begin{subfigure}{0.17\textwidth} 
    		\centering
    		\includegraphics[width=\textwidth]{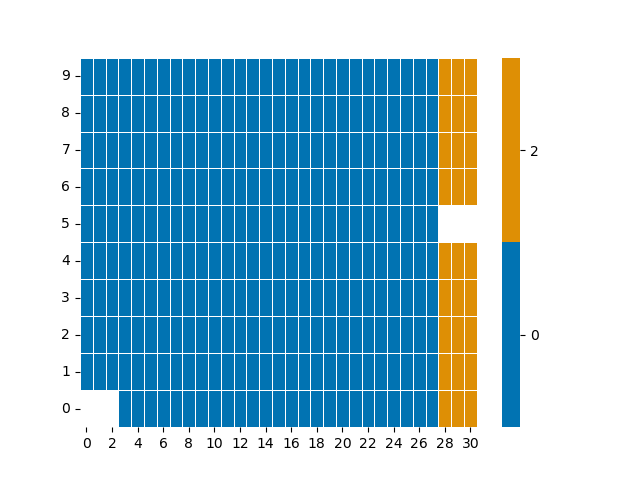}
    		\caption[Subfigure 11g]{\label{subfig:11g}}
	\end{subfigure}
	\begin{subfigure}{0.17\textwidth} 
    		\centering
    		\includegraphics[width=\textwidth]{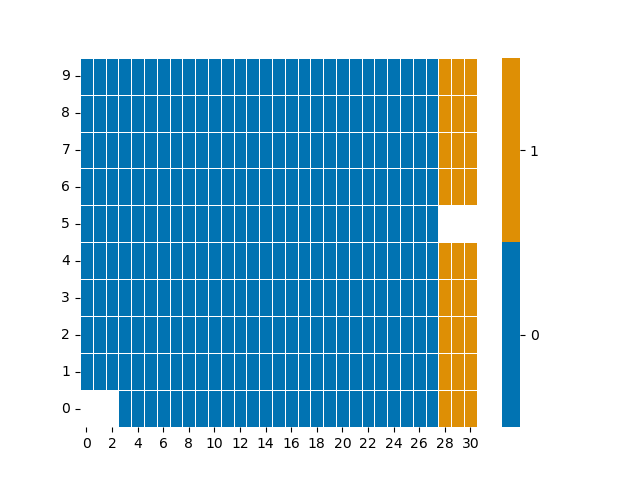}
    		\caption[Subfigure 11h]{\label{subfig:11h}}
	\end{subfigure}
	\begin{subfigure}{0.17\textwidth} 
    		\centering
    		\includegraphics[width=\textwidth]{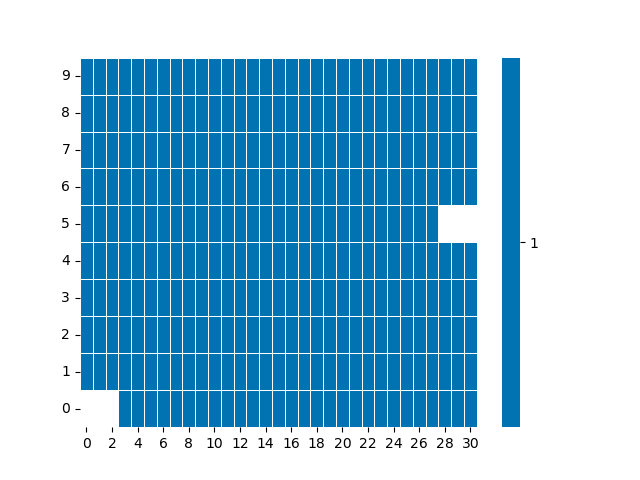}
    		\caption[Subfigure 11i]{\label{subfig:11i}}
	\end{subfigure}
    \caption[Fihure 11]{\label{fig:11}
    Results from quality control of a real-data example.
    Quality control based on the results of a linear model including block effects and IID (top)  or AR(1) (bottom) variance-covariance structure.
    Detected patches after identification (a, e), local authentication (b, f), and global authentication (c, g).
    Further merging of detected patches after final authentication with threshold 'anecdotal' to 'decisive' (d),
    smaller than or equal to 'moderate' (h),
    and greater than or equal to 'strong' (i).}
\end{figure}

The proposed method for automatic quality control confirms this observation.
It identifies twelve patches (see Figure \ref{subfig:11a}).
They are successively merged to ten and four patches through local and global authentication (see Figure \ref{subfig:11b} and \ref{subfig:11c}).
The patches are merged to two after final authentication (see Figure \ref{subfig:11d}).

Note that, in this example, mean and \textit{variance} stationarity was assumed; not mean and variance-\textit{covariance} stationarity.
We therefore restrict the quality control method to estimation of \gls{iid} variance-covariance structures.
Without this restriction, the method might detect only one residual patch if a stationary \gls{ar1} variance-covariance structure exists.
Still, this \gls{ar1} variance-covariance structure would not be adequately fitted by the \gls{iid} \gls{grf} from the analysis, leading to quality issues.
Furthermore, note that the proposed method is applied to marginal residuals (i.e., joint likelihood) while manual quality control is based on conditional residuals (i.e., easier visual detection).
For more information on the different types of residuals, refer to \citet{haslett2007three}.

\paragraph{Interpretation}
The spatial patterns in the residuals correspond to field effects that cannot be sufficiently estimated based on the experimental design and initial analysis.
Breeders assumed that field effects can be controlled using blocks and analysed under the hypothesis of otherwise constant disease pressure across the whole field (i.e., mean and \textit{variance} stationarity).
However, the phenotypic data indicates an autocorrelation structure (i.e., \textit{covariance}).
This is confirmed by the quality control method which divides the field along the direction of autocorrelation.

We assume that the direction of autocorrelation corresponds to the main wind direction.
Aphids attack the plants from the borders of the field, where they transmit the \gls{bydv} virus.
After several cycles of reproduction, they develop wings.
Aphids cannot fly in a targeted manner, but are carried by the wind to neighbouring plants where the cycle begins again \citep{jayasinghe2022effect}.
The disease is therefore not homogeneously spread, but depends on the entry points of aphids along the field border and their transmission by the wind.

\paragraph{Practical Implications}
This example highlights that quality issues may not only arise from \textit{nonstationary} itself, but also from insufficient model selection (e.g., using inappropriate \textit{type} of \textit{stationary} variance-covariance structure).
It is thus crucial to clarify which kind of stationarity is assumed for the analysis and which types of nonstationarity would consequently pose problem and need to be verified by the proposed method.
Using the current analysis would result in inaccurate estimations that depend not only on the actual resistance of varieties to \gls{bydv} but are confounded with the inhomogeneous disease pressure.
Practically, this implies that varieties may be selected, 
not because they are the most resistant ones, but because they were tested in parts of the field with comparably low disease pressure.

\paragraph{Recommendations for the analysis of the current trial}
To ensure accurate estimations and breeding decisions, we recommend using an \gls{ar1} variance-covariance structure for the analysis of the trial.
The resulting conditional residuals seem more \gls{iid} (see Figure \ref{subfig:10g}).
The automatic quality control method confirms this impression.
It detects one large and one very small residual patch after the second cycle of authentication (see Figure \ref{subfig:11g}).
The small residual patch corresponds to the rightmost columns of the trial which are separated by a row of missing values.
This might have influenced the estimation because the separation is in the direction of autocorrelation.
% , where the residuals indeed all lie much closer around zero compared to the rest of the trial (see Figure \ref{sub_disease_RCBD_AR_margres}) -> carful, visual should be done in cond res.
However, the strength of nonstationarity is not very high and final authentication detects two patches only for thresholds smaller or equal to 'moderate' (see Figure \ref{subfig:11h}) and otherwise merges them to two (see Figure \ref{subfig:11i}).

Adjusting the type of \gls{grf} is a rather simple solution and the proposed method would not have detected huge quality issues if model selection to choose the type of \gls{grf} had been performed in the first place.
In cases where stationarity cannot be assumed, the output of the proposed method (i.e., detected residual patches) could be directly used for analysis.
For example, if one of the residual patches is very small (e.g., in the real data example for thresholds below 'strong'), a naive approach may be to exclude the covered data from the analysis.
Another option may be to estimate the elements of the variance-covariance matrix that correspond to each residual patch separately and to correct the mean with respect to each residual patch.
This would directly account for deviations from mean and variance-covariance stationarity in the analysis.

\paragraph{Recommendations for the experimental design of future trials}
Apart from using the proposed method's output for analysis, it can be also used for improving the experimental design of future trials at the same location.
Disease trials are often conducted at the same location over several years since their success depends on local disease pressure.
Field effects may reoccur in a comparable manner, for instance, the main wind direction may remain consistent over years.

A first simple improvement for next year's design may be to rotate the blocking direction by 90 degrees to align with the main wind direction.
Additionally, more suitable experimental designs could be used, such as $\alpha$-design \citep{patterson1978block}.
They may provide greater flexibility in estimating gradients through the use of sub-blocks.
Here, the information about size and allocation of residual patches could be used to optimise size and allocation of sub-blocks.

\section{Discussion}
\label{sec:Discussion}

\paragraph{Performance of the proposed method}
We determine the proposed method to be well suited for quality control in agricultural field trials (i.e., detection of nonstationarity in grid-indexed data).
It estimates the number of residual patches well with only few cases of underestimation and thus reliably detects quality issues due to nonstationarity.
Furthermore, the similarity of detected to simulated patches is high.

Our simulations of mean and variance-covariance nonstationarity in two-di-
mensional grids are based on historical plant breeding field trials.
The method is capable of handling such realistic grid structures (e.g., missing data) and data (i.e., metric data from yield trials).
We also applied the method to a real dataset from plant breeding practice (i.e., ordinal data from disease trial), where the results align with the visual impression.

\paragraph{Relevance of the proposed method}
To the best of our knowledge, there are currently no other statistical methods available for detecting mean and variance-co-
variance nonstationarity in grid-indexed data, particularly in a fully automatic manner.
We believe that similar data structures and data types as the ones we tested exist in field trials of other domains (e.g., fertiliser or pesticide trials) and that the method therefore offers a broadly applicable solution for quality control in agricultural field trial practice.

\paragraph{Benefits of the proposed method}
The main advantage of the proposed method lies particularly in its automation.
Since the method does not require fine-tuning for individual trials, it can be efficiently applied across a large number of trials.
This allows to automatically flag those trials, where quality issues are detected (i.e., more than one residual patch is detected).
The method thus provides an initial time saving by reducing the number of trials that require closer inspection.
For flagged trials, information about the exact position and shape of detected residual patches provides additional guidance during manual inspection.

Another advantage of the proposed method is its transparency and intuitiveness.
The method mainly relies on partitioning the field with subsequent selection and merging of parts, with each step being visualisable.
This helps to confirm and interpret the method's output.
For instance, the severity of nonstationarity can be estimated by the number and size of residual patches.

\paragraph{Further potential of the proposed method}
In addition to verifying the assumption of stationarity, the method's output can be used to improve the analysis of conducted trials and the experimental design of future trials.
For example, the detected residual patches can be dropped from the analysis if they are small.
Or, one can directly account for them in the analysis by appropriately adapting expectation and variance-covariance matrix.
Information of position and size of detected residual patches may provide additional insights (e.g., biological, geological relationships).
In trials which are conducted over several years at the same location (e.g., disease trials), such information can be used to improve the experimental design of future trials.
For instance, the size and allocation of blocks from established designs (e.g., \gls{rcbd}, $\alpha$-design) can be optimised based on the size and allocation of residual patches.

We demonstrated these advantages and their relevance in practice when applying the proposed method to the real data example.
Apart from this, the real-data example illustrates the method's flexibility and motivates further generalisability to other models, particular to ordinal data analysis:
We analysed the scoring data assuming approximate normality, which is very common practice \citep{thoni1985auswertung, laidig2021breeding}.
The method is developed in a way, that makes it specific for two-dimensional grid structures (i.e., field trials) but keeps it flexible for different types of statistical processes (i.e., \gls{glm}).
More specifically, the steps of the method do not depend on the type of model that is fitted and as long as a score can be calculated for each step (i.e., to enable adequate model comparison or selection), the method is applicable.
For example, in Section \ref{sec:Real_Data_Illustration}, we showed that the method can be easily adapted to verify specific types of stationarity (e.g., mean and \textit{variance} stationarity or mean and variance-\textit{covariance} stationarity).
Similarly, we can adapt the method to \gls{glm}, although this would require substantial additional work and is suggested as a subject for future research.
In our work, we believe that changing the type of analysis of the real-data example would not have changed the validity of the general conclusions drawn from quality control (i.e., need to adjust for autocorrelation).

\paragraph{Direct nonstationary analysis versus the proposed method}
The proposed method is not meant to oppose nonstationary modeling techniques, which have been shown to be very helpful in the analysis of field trials \citep{rodriguez2018corr}.
Rather, it is intended as a preliminary step to verify if the assumption of stationarity is actually violated.
One can then manually assess if the result of the method aligns with the own impression from data acquisition and analysis.
We believe that leaving the final judgement to experts from the field rather than relying solely on statistical methods offers several benefits:

For example, residual patches may correspond to parts of a disease trial with no or very low disease pressure in comparison to the rest of the trial.
Including the data where varieties appear to be resistant solely due to absence of disease pressure may negatively influence breeding decisions, even when using nonstationary models.
Therefore, it may be preferable to exclude this part of the data completely to obtain more meaningful estimations for those varieties that remain estimable.

Furthermore, judging the severity of nonstationarity based on residual patches may help to prevent overfitting.
Using residual patches for nonstationary analysis can be seen as a form of post-blocking (i.e., using a more complex model than was considered when designing the experiment).
Increasing complexity of analysis post hoc, has been shown to be detrimental when taken too far \citep{gilmour2000post}.
Similarly, nonstationary modelling techniques may lead to overfitting, although the complexity is introduced via smooth adjustments rather than additional blocking.
In comparison to smooth nonstationary modelling techniques, the proposed method allows to control the level of complexity by regulating stringency of final authentication and manual selection of residual patches in terms of practical relevance.

\paragraph{Deep learning methods versus the proposed method}
As mentioned in Section \ref{sec:Related_Work}, the data structure of field trials is comparable to that of images.
Many methods for pattern recognition in images rely on the training of deep learning models \citep{lecun2015deep}.

Field trial data can, of course, be visualised as an image.
However, the pattern we want to detect is essentially of a statistical nature (i.e., realisations of a \gls{grf}).
It does not correspond to the shape of a physical object or organism that needs to be learned through training on image data.
Even though we do not know the exact appearance of the pattern in a specific dataset (i.e., parameters of the \gls{grf}), we know its underlying structure beforehand (i.e., types of \gls{grf}s).
Thus, although we could train a deep learning model for quality control in field trials, this is not necessary, and we therefore employ more classical statistical methods.

\paragraph{Computational improvements of the proposed method}
We determined binary tree as the best performing tree.
Since it is data-driven, its computational time is much higher than deterministic approaches.
We could improve the workflow of the whole method by directly combining tree indexation with Top-Down identification.
For instance, while the tree iteratively partitions the field, Top-Down algorithm could decide at each iteration which splits are accepted.
If a split of a part is not accepted, tree indexation would stop there and only continue with all other parts.
Thus, the tree does not need to be built until the plot level.
This would therefore decrease the computational time of the whole method.

\begin{figure}[H]
    \centering
	\begin{subfigure}{0.494\textwidth} 
    		\centering
    		\includegraphics[width=\textwidth]{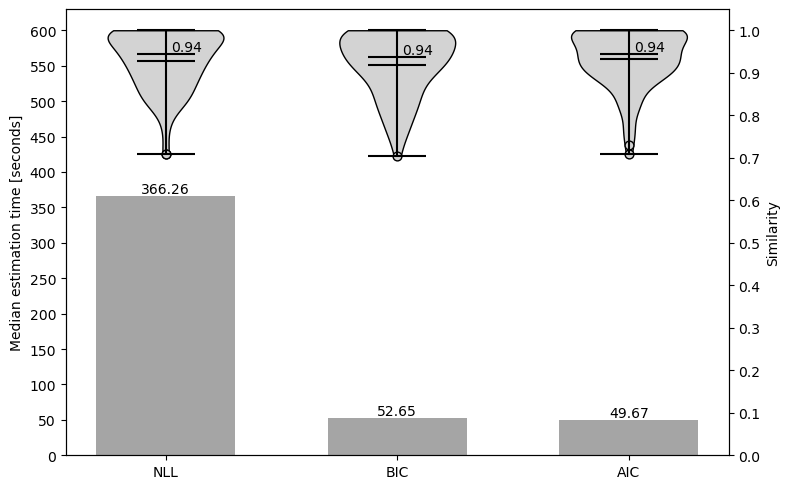}
    		\caption[Subfigure 12 a]
    		{\label{subfig:12a}}
	\end{subfigure}
	\begin{subfigure}{0.49\textwidth} 
    		\centering
    		\includegraphics[width=\textwidth]{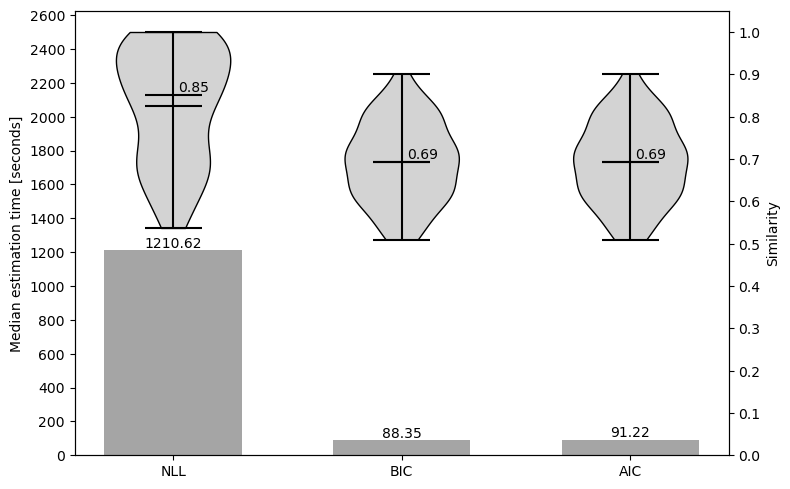}
    		\caption[Subfigure 12 b]
    		{\label{subfig:12b}}
	\end{subfigure}
    \caption[Figure 12]{\label{fig:12}
    Comparison in performance and estimation time for different scores.
    Performance specifically for binary tree indexation with precedent identification and authentication based on oracle algorithms (a) and for the entire method with binary tree indexation and identification using Top-Down algorithm. (b).
   }
\end{figure}

As score for all steps of the method, we use \gls{nll}.
Other scores may be more conventional.
As alternative scores, we tested \gls{bic} and \gls{aic}.
For tree indexation, they show similar performance as \gls{nll}, while requiring much less computational time (see Figure \ref{subfig:12a}).
However, the performance of the whole method based on \gls{bic} or \gls{aic} is comparably low (see Figure \ref{subfig:12b}) since it only detects one residual patch.
We therefore assume that \gls{bic} and \gls{aic} are not well suited for authentication.
In practical application, we advise to test building the tree with less time-consumping scores such as \gls{bic}.
For the subsequent steps, especially for authentication, \gls{nll} should be used.
This would decrease the overall computational time.
Nonetheless, it must be considered that \gls{nll} offers other advantages, that might compensate for its high computational time.
For instance, it is calculated based on cross validation, which makes it more robust to violations of model assumptions (e.g., normal distribution).
This may make the score more flexible for other data types such as ordinal disease scores, as in the real data example.

Apart from the predefined scores mentioned above, supervised machine learning could be applied to optimise the penalisation parameters.
This may improve the estimations of the overall method and provide a computationally more efficient alternative to for example \gls{nll}.
However, it would require some initial fine-tuning and may make the learned scores specific to the training dataset.

\paragraph{Improvement of the user-specific automation of the proposed method}
Without the final step of authentication, the method automatically detects location and shape of the residual patches well, but oversegments them.
This does not mean that the number of residual patches is not statistically meaningful.
For example, a more complex process (e.g., \gls{ar1}) may be well and even more parsimoniously described by several simpler processes (e.g., \gls{iid}).
However, if too many patches are detected, the output can become difficult for the user to interpret.
Therefore, we apply the last cycle of global authentication, where the user needs to set the threshold of stringency.

\begin{figure}[H]
	\centering
	\includegraphics[width=0.6\textwidth]{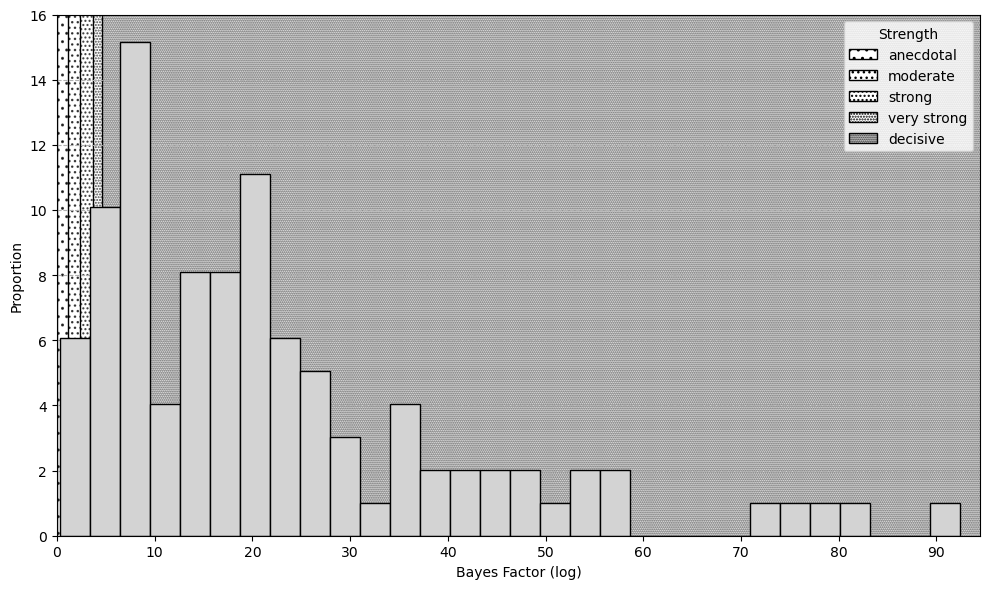}
 \caption[Figure 13]{\label{fig:13}
 	Strength of signal between the two simulated residual patches within each trial.
    }
\end{figure}

For our simulations we obtained only minor differences in the estimations for different thresholds (see Section \ref{subsec:Evaluation_authentication}).
This is due to the fact that the evidence for nonstationarity in our simulations (i.e., \gls{bf} calculated based on the simulated residual patches) mostly exceeds the threshold 'decisive' (see Figure \ref{fig:13}).
For the real data example, increasing thresholds only leads to differences for the analysis based on \gls{ar1} variance-covariance structure (i.e., detection of one residual patch instead of one large and one very small residual patch for thresholds greater than or equal to 'strong').

In practice, the appropriate threshold depends on the individual user's decision about the desired level of granularity in detecting nonstationarity.
We therefore recommend to apply the method to the specific dataset of the user, testing different thresholds for the final cycle of authentication.
Users can then manually evaluate which thresholds lead to meaningful results and apply the chosen threshold on a large scale for their quality control.

{\bf CRediT authorship contribution statement}\\
\textbf{Karen Wolf:}
Conceptualisation,
Methodology,
Software,
Formal analysis,
Writing – original draft,
Writing – review \& editing,
Visualisation.
\textbf{Pierre Fernique:}
Conceptualisation,
Methodology,
Software,
Formal analysis,
Data Curation,
Writing – review \& editing,
Supervision,
Project administration,
Funding acquisition.
\textbf{Hans-Peter Piepho:}
Writing – review \& editing,
Supervision,
Project administration.\\
\ \\
{\bf Declaration of competing interest}\\
The authors declare that they have no known competing financial interests or personal relationships that could have appeared to influence the work reported in this paper.\\
\ \\
{\bf Data availablility}\\
The authors do not have permission to share data or code.\\
\ \\
{\bf Acknowledgements} \\
This work was supported by Limagrain Europe.
The authors sincerely thank Fabien Le Couviour for his assistance in providing a suitable real-data example.
We are also grateful to Guillaume Noyel for his valuable guidance in identifying an appropriate journal for the submission of this manuscript.

%% \section{App}
%% \label{app1}

%%       \citet{<label>} ==> Lamport (1994)
%%       \citep{<label>} ==> (Lamport, 1994)
%%
% Example citation, See \citet{lamport94}.
% \ \\

\begin{thebibliography}{00}

% \bibitem[Lamport(1994)]{lamport94}
%   \textit{\LaTeX: a document preparation system},
%   Addison Wesley, Massachusetts,

\bibitem[Qaim(2020)]{qaim2020role}
M. Qaim,
\textit{Role of new plant breeding technologies for food security and sustainable agricultural development},
Appl. Econ. Perspec. and Policy,
Wiley Online Library
42(2)
(2020)
129--150.
https://doi.org/10.1002/aepp.13044.

\bibitem[M{\"o}hring and Piepho(2009)]{mohring2009comparison}
J. M{\"o}hring, H.P. Piepho,
\textit{Comparison of weighting in two-stage analysis of plant breeding trials},
Crop Sci.,
Wiley Online Library
49(6)
(2009)
1977--1988.
https://doi.org/10.2135/cropsci2009.02.0083.

\bibitem[Piepho et al.(2012)]{piepho2012stage}
H.-P. Piepho, J. M{\"o}hring, J., T. Schulz-Streeck, J.O. Ogutu,
\textit{A stage-wise approach for the analysis of multi-environment trials},
Biom. J.,
Wiley Online Library
54(6)
(2012)
844--860.
https://doi.org/10.1002/bimj.201100219.
 
\bibitem[Kumar et al.(2024)]{kumar2024advances}
R. Kumar, S.P. Das, B.U. Choudhury, A. Kumar, N.R. Prakash, R. Verma, M. Chakraborti, A.G. Devi, B. Bhattacharjee, R. Das, B. Das, H.L. Devi, B.Das, S. Rawat, V.K. Mishra,
\textit{Advances in genomic tools for plant breeding: Harnessing DNA molecular markers, genomic selection, and genome editing},
Biol. Res.,
Springer
57(1)
(2024)
80.
https://doi.org/10.1186/s40659-024-00562-6.

\bibitem[Cressie(1993)]{cressie1993statistics}
N.A.C. Cressie,
\textit{Statistics for spatial data (revised edition)},
John Wiley \& Sons,
New York,
1993.
https://doi.org/10.1002/9781119115151.

\bibitem[Mao et al.(2020)]{mao2020adjusting}
X. Mao, S. Dutta, R.K.W. Wong, D. Nettleton,
\textit{Adjusting for Spatial Effects in Genomic Prediction},
JABES,
Springer
25(4)
(2020)
699--718.
https://doi.org/10.1007/s13253-020-00396-1.

\bibitem[John and Williams(1995)]{john1995cyclic}
J.A. John, E.R. Williams,
\textit{Cyclic and computer generated designs},
2nd edition
Chapman and Hall,
New York
1996.
https://doi.org/10.1201/b15075.

\bibitem[Bailey(2008)]{bailey2008design}
R.A.Bailey,
\textit{Design of comparative experiments},
Volume 25
Cambridge University Press,
Cambridge
2008.
https://doi.org/10.1017/CBO9780511611483.

\bibitem[Butler et al.(2017)]{butler2017asreml}
D.G. Butler, B.R. Cullis, A.R. Gilmour, B.J. Gogel, R. Thompson,
\textit{ASReml-R reference manual version 4},
VSN Int. Ltd,
Hemel Hempstead,
UK,
2017.

\bibitem[Pearce(1995)]{pearce1995some}
Pearce, S.C.,
\textit{Some design problems in crop experimentation. I. The use of blocks},
Exp. Agric.,
Cambridge University Press,
31(2)
(1995)
191--204.
https://doi.org/10.1017/S0014479700025278.

\bibitem[Fisher(1935)]{fisher1935design}
R.A. Fisher,
\textit{The Design of Experiments},
Oliver \& Boyd,
Edinburgh,
1935.

\bibitem[Yates(1936)]{yates1936new}
F. Yates,
\textit{A new method of arranging variety trials involving a large number of varieties},
The J. of Agric. Sci.,
Cambridge University Press,
26(3)
(1936)
424--455.
https://doi.org/10.1017/S0021859600022760.

\bibitem[Patterson et al.(1978)]{patterson1978block}
H.D. Patterson, E.R. Williams, E.A. Hunter,
\textit{Block designs for variety trials},
The J. of Agric. Sci.,
Cambridge University Press
90(2)
(1978)
395--400.
https://doi.org/10.1017/S0021859600055507.

\bibitem[Kempton(1984)]{kempton1984design}
Kempton, R.A.,
\textit{The design and analysis of unreplicated field trials},
Vor. f{\"u}r Pflanzenz{\"u}cht.,
Plant Breeding Institute,
Cambridge,
UK
7
(1984)
219--242.

\bibitem[Wang et al.(2024)]{wang2024decomposition}
H. Wang, L. Al Tarawneh, C. Cheng, Y. Jin,
\textit{A decomposition-guided mechanism for nonstationary time series forecasting},
AIP Adv.,
AIP Publishing
14(1)
(2024).
https://doi.org/10.1063/5.0153647.

\bibitem[Osborne and Suárez-Seoane(2002)]{osborne2002should}
P.E. Osborne, S. Suárez-Seoane,
\textit{Should data be partitioned spatially before building large-scale distribution models?},
Ecol. Model.,
Elsevier
157(2-3)
(2002)
249--259.
https://doi.org/10.1016/S0304-3800(02)00198-9.

\bibitem[Mercer and Hall(1911)]{mercer1911experimental}
W.B. Mercer, A.D. Hall,
\textit{The experimental error of field trials},
The J. of Agric. Sci.,
Cambridge University Press
4(2)
(1911)
107--132.
https://doi.org/10.1017/S002185960000160X.

\bibitem[Rodr{\'\i}guez-\'Alvarez et al.(2018)]{rodriguez2018corr}
M.X. Rodr{\'\i}guez-\'Alvarez, M.P. Boer, F.A. van Eeuwijk, P.H.C. Eilers,
\textit{Correcting for spatial heterogeneity in plant breeding experiments with P-splines},
Spat. Stat.,
Elsevier
23
(2018)
52--71.
https://doi.org/10.1016/j.spasta.2017.10.003.

\bibitem[Dreesman and Tutz(2001)]{dreesman2001non}
J.M. Dreesman, G. Tutz,
\textit{Non-stationary conditional models for spatial data based on varying coefficients},
J. of the R. Stat. Soc.: Ser. D (The Stat.),
Wiley Online Library
50(1)
(2001)
1--15.
https://doi.org/10.1111/1467-9884.00256.

\bibitem[Brunsdon et al.(1996)]{brunsdon1996geographically}
C. Brunsdon, A.S. Fotheringham, M.E. Charlton,
\textit{Geographically weighted regression: A method for exploring spatial nonstationarity},
Geogr. Anal.,
Wiley Online Library
28(4)
(1996)
281--298.
https://doi.org/10.1111/j.1538-4632.1996.tb00936.x.

\bibitem[Priestley(1965)]{priestley1965evolutionary}
M.B. Priestley,
\textit{Evolutionary spectra and non-stationary processes},
J. of the R. Stat. Soc.: S. B (Methodol.),
Wiley Online Library,
27(2)
(1965)
204--229.
https://doi.org/10.1111/j.2517-6161.1965.tb01488.x.

\bibitem[Dahlhaus(2000)]{dahlhaus2000likelihood}
R. Dahlhaus,
\textit{A likelihood approximation for locally stationary processes},
The Ann. of Stat.,
Institute of Mathematical Statistics
28(6)
(2000)
1762--1794.
https://doi.org/10.1214/aos/1015957480.

\bibitem[Kim et al.(2005)]{kim2005analyzing}
H.-M. Kim, B.K. Mallick, C.C. Holmes,
\textit{Analyzing nonstationary spatial data using piecewise Gaussian processes},
J. of the Am. Stat.l Assoc.,
Taylor \& Francis
100(470)
(2005)
653--668.
https://doi.org/10.1198/016214504000002014.

\bibitem[Tzeng et al.(2024)]{tzeng2024assessing}
S. Tzeng, B.Y. Chen,H.-C.Huang,
\textit{Assessing spatial stationarity and segmenting spatial processes into stationary components},
JABES,
Springer
29(2)
(2024)
301--319.
https://doi.org/10.1007/s13253-023-00588-5.

\bibitem[Corwin and Lesch(2010)]{corwin2010delineating}
D.L. Corwin, S.M.Lesch,
\textit{Delineating site-specific management units with proximal sensors},
Geostat. Appl. for Precis. Agric.,
Springer
(2010)
139--165.
https://doi.org/10.1007/978-90-481-9133-8\_6.

\bibitem[Rakshit et al.(2020)]{rakshit2020novel}
S. Rakshit, A. Baddeley, K. Stefanova, K. Reeves, K. Chen, Z. Cao, F. Evans, M. Gibberd,
\textit{Novel approach to the analysis of spatially-varying treatment effects in on-farm experiments},
Field Crops Res.,
Elsevier
255
(2020)
107783.
https://doi.org/10.1016/j.fcr.2020.107783.

\bibitem[Lacasa(2023)]{lacasa2023bayesian}
J. Lacasa,
\textit{A Bayesian approach for estimating and checking block designs in agricultural experiments},
Master thesis,
Kansas State University,
Manhattan,
Kansas,
2023.

\bibitem[Guinness and Fuentes(2015)]{guinness2015likelihood}
J. Guinness, M. Fuentes,
\textit{Likelihood approximations for big nonstationary spatial temporal lattice data},
Stat. Sinica,
JSTOR
25(1)
(2015)
329--349.
https://www.jstor.org/stable/24311019.

\bibitem[Hunter and Steiglitz(1979)]{hunter1979operations}
G.M. Hunter, K. Steiglitz,
\textit{Operations on images using quad trees},
IEEE Trans. on Pattern Anal. and Mach. Intell.,
IEEE
(2)
(1979)
145--153.
10.1109/TPAMI.1979.4766900.

\bibitem[Gramacy and Lee(2008)]{gramacy2008bayesian}
R.B. Gramacy, H.K.H. Lee,
\textit{Bayesian treed gaussian process models with an application to computer modeling},
J. of the Am. Stat. Assoc.,
Taylor \& Francis
103(483)
(2008)
1119--1130.
https://doi.org/10.1198/016214508000000689.


\bibitem[Krass and Raftery (1995)]{krass1995bayes}
R.E. Krass, A.E. Raftery,
\textit{Bayes factors},
J. of the Am. Stat. Assoc.,
Taylor \& Francis
90(430)
(1995)
773–795.
https://doi.org/10.1080/01621459.1995.10476572.

\bibitem[Lauritzen(1996)]{lauritzen1996graphical}
S.L. Lauritzen,
\textit{Graphical models},
Volume 17,
Clarendon Press,
Oxford, UK,
1996.

\bibitem[Lloyd(1982)]{lloyd1982least}
S. Lloyd,
\textit{Least squares quantization in PCM},
IEEE Trans. on Inf. Theory,
IEEE
28(2)
(1982)
129--137.
https://doi.org/10.1109/TIT.1982.1056489.

\bibitem[Russell(1978)]{russell1978plant}
E.G. Russell,
\textit{Plant breeding for pest and disease resistance: studies in the agricultural and food sciences},
Butterworths,
London
1978.

\bibitem[Hickey et al.(2012)]{hickey2012rapid}
L.T. Hickey, P.M. Wilkinson, C.R. Knight, I.D. Godwin, O.Y. Kravchuk, E.A.B. Aitken, U.K. Bansal, H.S. Bariana, I.H. De Lacy, M.J. Dieters,
\textit{Rapid phenotyping for adult-plant resistance to stripe rust in wheat.},
Pl. Breed.,
Wiley Online Library
131(1)
(2012)
54--61.
https://doi.org/10.1111/j.1439-0523.2011.01925.x.

\bibitem[Miller and Rasochov{\'a}(1997)]{miller1997barley}
W.A. Miller, L. Rasochov{\'a},
\textit{Barley yellow dwarf viruses},
Annu. rev. of phytopathol.,
Annual Reviews
35(1)
(1997)
167--190.
https://doi.org/10.1146/annurev.phyto.35.1.167.

\bibitem[Th{\"o}ni(1985)]{thoni1985auswertung}
H. Th{\"o}ni,
\textit{Auswertung von Bonituren: Ein empirischer Methodenvergleich},
EDV in Med. und Biol.,
Gustav Fischer Verlag GK
16(3)
(1985)
108--114.

\bibitem[Laidig et al.(2021)]{laidig2021breeding}
F. Laidig, T. Feike, S. Hadasch, D. Rentel, B. Klocke, T. Miedaner, H. P. Piepho,
\textit{Breeding progress of disease resistance and impact of disease severity under natural infections in winter wheat variety trials},
Theor. and Appl. Genet.,
Springer
134
(2021)
1281--1302.
https://doi.org/10.1007/s00122-020-03728-4.

\bibitem[Haslett and Haslett (2007)]{haslett2007three}
J. Haslett, S.J. Haslett,
\textit{The three basic types of residuals for a linear model},
Int. Stat. Rev.,
Wiley Online Library
75(1)
(2007)
1--24.
https://doi.org/10.1111/j.1751-5823.2006.00001.x.

\bibitem[Jayasinghe et al.(2022)]{jayasinghe2022effect}
W.H. Jayasinghe, M.S. Akhter, K. Nakahara, M.N. Maruthi,
\textit{Effect of aphid biology and morphology on plant virus transmission},
Pest Manag. Sci.,
Wiley Online Library
78(2)
(2022)
416--427.
https://doi.org/10.1002/ps.6629.

\bibitem[Gilmour(2000)]{gilmour2000post}
A.R. Gilmour,
\textit{Post blocking gone too far! Recovery of information and spatial analysis in field experiments},
Biom.,
Wiley Online Library
56(3)
(2000)
944--945.
https://doi.org/10.1111/j.0006-341X.2000.944\_1.x.

\bibitem[LeCun et al.(2015)]{lecun2015deep}
Y. LeCun, Y. Bengio, G. Hinton,
\textit{Deep learning},
Nat.,
521
(2015)
436–-444.
https://doi.org/10.1038/nature14539

\end{thebibliography}


\bibliographystyle{plain}

\ \\
\textbf{Karen Wolf}
received her M.S. degrees in Plant Breeding and Statistics, both awarded by the University of Goettingen in 2023.
Currently, she is pursuing her Ph.D. degree at the University of Hohenheim.
Her research interests include field trial design and analysis.\\
\ \\
\textbf{Pierre Fernique}
received his Ph.D. in Applied Statistics from the University of Montpellier.
He is currently working as a researcher at the French plant breeding company Limagrain.
His research interests include optimising the design and analysis of field trials.\\
\ \\
\textbf{Hans-Peter Piepho}
has a Ph.D. in Plant Breeding and is a Professor of Bioststistics at the University of Hohenheim.
His research interests include linear mixed models and optimal experimental design for plant breeding and variety testing.
%\printglossary[type=\acronymtype, title={List of Acronyms}]
\printglossary[type=\acronymtype]

\end{document}